\documentclass[apj]{emulateapj}
\usepackage[]{natbib}
\usepackage{graphicx, amsmath, multirow}
\usepackage{hyperref,breakurl}
\bibpunct{(}{)}{,}{a}{}{,}

\begin{document}
\def\bullet{\object{1E0657$-$56}}
\def\bbullet{\object{MACS~J0025.4$-$1222}}
\def\macsos{MACS~J0717$+$3745}
\def\macsa{MACS~J0416.1$-$2403}
\def\macsb{MACS~J1149.5$+$2223}
\def\macsc{RXC~J2248.7$-$4431}
\def\HST{{\it HST}}
\def\Spitzer{\it Spitzer}
\def\arcsecf{\!\!^{\prime\prime}}
\def\arcminf{\!\!^{\prime}}
\def\diff{\mathrm{d}}
\def\ngx{N_{\mathrm{x}}}
\def\ngy{N_{\mathrm{y}}}
\def\eck#1{\left\lbrack #1 \right\rbrack}
\def\eckk#1{\bigl[ #1 \bigr]}
\def\round#1{\left( #1 \right)}
\def\abs#1{\left\vert #1 \right\vert}
\def\wave#1{\left\lbrace #1 \right\rbrace}
\def\ave#1{\left\langle #1 \right\rangle}
\def\kms{{\rm \:km\:s}^{-1}}
\def\dds{D_{\mathrm{ds}}}
\def\dd{D_{\mathrm{d}}}
\def\ds{D_{\mathrm{s}}}
\def\cs{\mbox{cm}^2\mbox{g}^{-1}}
\def\magz{m_{\rm z}}
\def\V{\rm{V_{\rm 606}}}
\def\ii{\rm{i_{\rm 775W}}} 
\def\iii{\rm{I_{\rm 814W}}} 
\def\z{\rm{z_{\rm 850LP}}}
\def\J{\rm{J_{\rm 110W}}}
\def\H{\rm{H_{\rm 160W}}}
\def\chone{3.6\:\mu\rm{m}}
\def\chtwo{4.5\:\mu\rm{m}}
\def\cunit{\mbox{erg}/\mbox{s}/\mbox{cm}^2/\mbox{\AA}}
\def\lunit{\mbox{erg}/\mbox{s}/\mbox{cm}^2}
\def\ct{[CII]}
\def\lya{Lyman-$\alpha$}
\newcommand{\hst}{\it HST}
\newcommand{\galfit}{\texttt{GALFIT}}    
\newcommand{\mopex}{\texttt{mopex}}      
\newcommand{\surfsup}{SURFS UP}         
\newcommand{\sex}{\texttt{SExtractor}}   
\newcommand{\sersic}{S\'ersic}           
\newcommand{\pygfit}{\texttt{PyGFIT}}    
\newcommand{\lephare}{\texttt{Le Phare}} 
\newcommand{\eqn}[1]{equation~(\ref{#1})}
\newcommand{\fig}[1]{Figure~\ref{#1}}
\newcommand{\tab}[1]{Table~\ref{#1}}

\newcommand{\nframes}{2100}              
\newcommand{\naors}{5}                   
\newcommand{\nclust}{10}                 
\newcommand{\ngal}{10}                   

\title{Spitzer UltRa Faint SUrvey Program ({\surfsup}) I: An Overview \altaffilmark{*}}
\altaffiltext{*}{These observations are associated with
programs {\Spitzer} \#90009, 60034, 00083, 50610, 03550, 40593, and {\hst} \# GO10200, GO10863, GO11099, and GO11591. Furthermore based on ESO Large Program ID 181.A-0485.}
\shorttitle{}
\author{Maru\v{s}a Brada\v{c}\altaffilmark{1},
Russell Ryan\altaffilmark{2},
Stefano Casertano\altaffilmark{2},
Kuang-Han Huang\altaffilmark{1},
Brian \ C. Lemaux\altaffilmark{4},
Tim Schrabback\altaffilmark{3},
Anthony \ H. Gonzalez\altaffilmark{5},
Steve Allen\altaffilmark{6},
Benjamin Cain\altaffilmark{1},
Mike Gladders\altaffilmark{7},
Nicholas Hall\altaffilmark{1},
Hendrik Hildebrandt\altaffilmark{3},
Joannah Hinz\altaffilmark{8},
Anja von der Linden\altaffilmark{6,9},
Lori Lubin\altaffilmark{1},
Tommaso Treu\altaffilmark{10,11,x},
Dennis\ Zaritsky\altaffilmark{8}}
\shortauthors{Brada\v{c} et al.}
\altaffiltext{1}{Department of Physics, University of California, Davis, CA 95616, USA}
\altaffiltext{2}{Space Telescope Science Institute, 3700 San Martin
  Drive, Baltimore, MD 21218, USA}
\altaffiltext{3}{Argelander-Institut f\"{u}r Astronomie, Auf dem H\"{u}gel 71, D-53121 Bonn, Germany}
\altaffiltext{4}{Aix Marseille Universit\'{e}, CNRS, LAM (Laboratoire d'Astrophysique de Marseille) UMR 7326, 13388, Marseille, France}
\altaffiltext{5}{Department of Astronomy, University of Florida, 211 Bryant Space Science Center, Gainesville, FL 32611, USA}
\altaffiltext{6}{Kavli Institute for Particle Astrophysics and Cosmology, Stanford University, 382 Via Pueblo Mall, Stanford, CA 94305-4060, USA}
\altaffiltext{7}{The University of Chicago,The Kavli Institute for Cosmological Physics,
933 East 56th Street,
Chicago, IL 60637, USA}
\altaffiltext{8}{Steward Observatory, University of Arizona, 933 N Cherry Ave., Tucson, AZ 85721, USA}
\altaffiltext{9}{Dark Cosmology Centre, Niels Bohr Institute, University of Copenhagen, Juliane Maries Vej 30, 2100 Copenhagen {\O}, Denmark}
\altaffiltext{10}{Department of Physics, University of California, Santa Barbara, CA 93106, USA}
\altaffiltext{11}{KITP, Kohn Hall, UC Santa Barbara, Santa Barbara CA 93106-4030} 
\altaffiltext{x}{Packard Fellow}
\email{marusa@physics.ucdavis.edu}


\begin{abstract} {\surfsup} is a joint {\Spitzer} and {\hst}
  Exploration Science program using 10 galaxy clusters as cosmic
  telescopes to study $z\gtrsim 7$ galaxies at intrinsically lower
  luminosities, enabled by gravitational lensing, than blank field
  surveys of the same exposure time. Our main goal is to measure
  stellar masses and ages of these galaxies, which are the most likely
  sources of the ionizing photons that drive reionization. Accurate
  knowledge of the star formation density and star formation history
  at this epoch is necessary to determine whether these galaxies
  indeed reionized the universe. Determination of the stellar masses
  and ages requires measuring rest frame optical light, which only
  {\Spitzer} can probe for sources at $z\gtrsim 7$, for a large enough
  sample of typical galaxies. Our program consists of 550 hours of
  Spitzer/IRAC imaging covering 10 galaxy clusters with very
  well-known mass distributions, making them extremely precise cosmic
  telescopes. We combine our data with archival observations to obtain
  mosaics with $\sim 30$ hours exposure time in both ${\chone}$ and
  ${\chtwo}$ in the central $4\arcmin \times 4\arcmin$ field and $\sim
  15$ hours in the flanking fields. This results in 3-$\sigma$
  sensitivity limits of $\sim26.6$ and $\sim 26.2$\,AB magnitudes for
  the central field in the IRAC 3.6 and ${\chtwo}$ bands,
  respectively. To illustrate the survey strategy and characteristics
  we introduce the sample, present the details of the data reduction
  and demonstrate that these data are sufficient for in-depth studies
  of $z\gtrsim 7$ sources (using a $z=9.5$ galaxy behind {\macsb} as
  an example). For the first cluster of the survey (the Bullet
  Cluster) we have released all high-level data mosaics and IRAC
  empirical PSF models. In the future we plan to release these data
  products for the entire survey.  
\end{abstract}

\section{Introduction}
\label{sec:intro}

{\surfsup} ({\Spitzer} UltRa Faint SUrvey Program: Cluster Lensing and
{\Spitzer} Extreme Imaging Reaching Out to $z\gtrsim 7$, \#90009 PI
Brada\v{c}, co-PI Schrabback) is a joint {\Spitzer} and {\hst}
Exploration Science program. It was designed to image 10 galaxy
cluster fields to extreme depths with {\Spitzer} ${\chone}$ and
${\chtwo}$ bands for 550 hours total. It also includes 13 prime and 13
parallel orbits of {\hst} time for one of the clusters which did not
have deep WFC3-IR and optical {\hst} data (RCS2-2327.4$-$0204; the
rest of the targets have {\hst} data available).  Together with the
archival data, each field has been or will be imaged with {\Spitzer}
for $>100\mbox{ks}$ (28 hours) per band. Such depths have only been
achieved previously with {\Spitzer} observations of the Ultra Deep
Field UDF \citep{labbe12, vgonzalez10, labbe10}, GOODS (40 hours per field, see below and e.g., \citealp{oesch13}) and CANDLES
  through the S-CANDELS program (P.I. G. Fazio; five CANDELS fields to
  50 hours depth with IRAC). In the near future, SPLASH Survey
  (Spitzer Large Area Survey with Hyper-Suprime-Cam, PI Capak,
  \#90042) will provide 2475h of Spitzer observing over two
  $1.8\deg^2$ fields (COSMOS and SXDS); delivering depths of $\sim 10$
  hours per pointing. Compared to these studies, {\surfsup} has the advantage of studying intrinsically lower
    luminosities, enabled by gravitational lensing, than blank field
    surveys of the same exposure time and has been designed to address the two main science goals described below.

\subsection{Star formation at $z\gtrsim 7$}
The epoch of reionization marked the end of the so-called ``Dark
Ages'' and signified the transformation of the universe from opaque to
transparent. Yet the details of this important transition period are
still poorly understood.  A compelling but most likely overly
simplistic suggestion is that star-forming galaxies at $z\gtrsim 7$
are solely responsible for reionization. The ability of sources to
reionize the universe depends in part upon their co-moving star
formation rate density $\rho_{\rm SFR}$ and star formation history at
high redshift (for reviews, see \citealp{fan06, robertson10,
  loeb12}). The advent of Wide Field Camera 3 (WFC3) on {\hst} detects
these galaxies at rest-frame UV wavelengths (e.g., \citealp{ellis13,
  bouwens13,schenker13}), while {\Spitzer} observations allow us to
trace the rest-frame continuum emission redward of $4000\mbox{\AA}$
(e.g., \citealp{labbe12}). Rest-frame UV and rest-frame
$4000\mbox{\AA}$ data trace two basic properties of stellar
populations; the instantaneous star formation rate (SFR) dominated by
younger stars and the integrated history of the older population,
respectively \citep{madau99}. The stellar masses allow us to determine
the SFR density at $z \gtrsim 7$, which can be compared to the SFR
density needed for these sources to reionize the universe (for certain
choices of escape and clumping factors, \citealp{madau99, robertson13,
  stark13}).
 
{\surfsup} has the advantage that by using deep observations of 10
independent sight lines sample variance is reduced compared to e.g.,
UDF. Clusters of galaxies, when used as cosmic telescopes, allow us to
probe deeper due to high magnification and {\surfsup} targets are
among the largest galaxy clusters known and were chosen for their
extreme lensing strength. This program therefore allows us to push the
intrinsic luminosity limits further than the UDF and study
representative galaxies at $z\sim 7$ and 8. For example, clusters that
are part of this survey have typical magnifications of $\mu \gtrsim
5$, which effectively increases the exposure time by $\sim \mu^2$. For
the $z\sim 9.5$ galaxy reported below the {\it intrinsic} (corrected
for lensing) measured magnitudes in IRAC are $28.6^{+0.9}_{-0.8}$ in
$\chone$ and $27.9^{+0.6}_{-0.4}$ in $\chtwo$, compared to 5-$\sigma$
limiting magnitudes reported by \citep{oesch13} in GOODS-N of 27.0 and
26.7, respectively.

One concern, however, when using gravitational lensing is that lensing
magnification decreases the effective observing field (as it
``enlarges'' sources and their separations on the sky).  This loss in
sky area is more than compensated for by the steep luminosity function
(effective slope $>2$) at the magnitudes that we probe
\citep{bouwens12, bradley12}. A second concern is that we need to know
the magnification (including errors) of our cosmic telescopes to
convert the observed number counts and stellar masses into their
intrinsic values. As shown by \citet{bradac09}, the magnification of
well-studied clusters, needed for such conversion, can be constrained using information on
distortion and shifts of the background sources to sufficient
accuracy. In summary, (1) in the regimes where the luminosity function
is steep (effective slope $>2$, which is true at the magnitudes that
we probe) number counts are increased compared to observations in a
blank field (i.e., many somewhat fainter galaxies become accessible
because of the foreground lens), and (2) magnification errors amount
to a smaller error than sample variance when determining the
luminosity function at $z\sim 7$. Another advantage of gravitational
lensing is that lensed galaxies are often enlarged, easing
identification (gravitational lensing magnifies solid angles while
preserving colors and surface brightness).

 The first demonstration of
an established stellar population at high redshift ($z\gtrsim 6$) was
accomplished using {\Spitzer} data of the strongly-lensed $z\sim 6.8$
galaxy behind Abell 2218 \citep{egami05, kneib04}. Detections at
${\chone}$ and ${\chtwo}$ allowed the construction of the galaxy's
spectral energy distributions (SED) and measurement of the stellar
properties. 
The SED has a significant rest-frame $4000\mbox{\AA}$ break and
therefore indicates that a mature stellar population is already in
place at such a high redshift \citep{egami05}. These measurements were
made possible due to large magnification factors ($\sim 25$). When
observing gravitationally magnified objects, {\Spitzer}/IRAC imaging
enables us to study stellar populations of the highest redshift
galaxies (e.g., see \citealp{zheng12} for a $z\sim 9.5$ galaxy detected
by {\Spitzer}; \citealp{smit13} for detections at $z = 6.6-7.0$).

Considerable investment has recently been made in observing galaxy
clusters with HST. The Cluster Lensing And Supernova survey with
Hubble (CLASH; \citealp{postman12}) delivered observations of 25
clusters and {\hst}-GO-11591 (PI Kneib) observed an additional 9
clusters.  Future high redshift exploration will be advanced by the
{\hst} Frontier Field
HFF\footnote{\url{http://www.stsci.edu/hst/campaigns/frontier-fields/}}
program, a program involving six deep fields centered on strong
lensing galaxy clusters in parallel with six deep “blank fields” (PIs
Mountain, Lotz). Very deep {\Spitzer} data are an excellent complement
to deep {\hst} data, which CLASH does not provide. The typical
integration times for CLASH clusters prior to {\surfsup} range from
$\sim 3.5$ hours per IRAC band from the ICLASH program (\#80168: PI
Bouwens, \citealp{bouwens12c}) to $\sim 5$ hours per IRAC band from
the Spitzer IRAC Lensing Survey program (\#60034: PI Egami).
{\surfsup} provides the greater depth and coverage needed in 10 strong
lensing clusters specifically chosen for their high lensing strength
(see below, 2 of them are part of HFF). The {\Spitzer} campaign covering the HFF will provide
similar depth for at least additional 2 clusters. In summary, {\Spitzer} plays a unique
role in the investigation of stellar ages and masses of $z\gtrsim 7$
galaxies. IRAC ${\chone}$ and ${\chtwo}$ observation probe rest-frame
optical wavelengths ($\sim 0.5\mu\mbox{m}$) which are the only
available data redward of rest-frame $4000\mbox{\AA}$ for these
sources and hence can probe presence of evolved stellar populations
for a large number of distant sources.

\subsection{Evolution of Stellar Mass Function in Galaxy Clusters}
With the {\surfsup} observations, we will also be able to probe the
stellar mass function of $z\sim 0.3-0.7$ members of our cluster sample
to depths of $<10^8$ M$_{\odot}$ (or 0.005 L$^*$) for an elliptical
galaxy at the highest cluster redshift we probe ($z=0.7$). This depth
far exceeds the current limits from studies of other high-redshift
clusters (e.g., \citealp{andreon06b, patel09, demarco10, vanderburg13}) and is
comparable to the deepest observations of local clusters, such as the
Coma cluster and the Shapley Supercluster
\citep{terlevich01,merluzzi10}. Furthermore, it is comparable to the
state-of the art stellar mass surveys at low redshifts; e.g., the
Galaxy And Mass Assembly (GAMA) survey \citep{taylor11, baldry12},
which has limits of $M^{*}\sim 0.5-1 \times 10^8$ M$_{\odot}$ at a
median redshift of $z = 0.05$. Previous optical/near-IR observations
of galaxy clusters at $z \sim 0.8$ suggested a deficit of faint, red
galaxies in the cluster red sequence (RS) as indicated by the
color-magnitude diagram (CMD) and the RS luminosity function (e.g.,
\citealp{delucia04,delucia07, tanaka05,rudnick09, gilbank10,
  lemaux12}). However, other authors find no such deficit (e.g.,
\citealp{depropris13, crawford09,andreon06, andreon08}). Because
rest-frame optical luminosities can be strongly affected by current
star-formation, color--stellar mass plots can look substantially
different than CMDs (e.g., \citealp{lemaux12}).
As a result, the stellar mass function and the processes governing its
evolution are the most
physical and accurate way to trace evolution in the cluster galaxy
population. So far, there has been little observed evolution in the
cluster stellar mass function; however, published results have only
probed down to $\sim 10^{10}$ M$_{\odot}$ (e.g., \citealp{bell04,
  demarco10, vulcani13}).

Because evolution is accelerated in overdense environments (e.g.,
\citealp{tanaka08}), it is essential to probe to lower stellar mass
limits in the cluster cores to get a complete picture of galaxy
evolution in these regions. {\surfsup} will achieve that by
  making a complete census of star forming cluster galaxies down to
  stellar masses of $10^8$ M$_{\odot}$ (or 0.005 L$^*$). Combined with
  our optical and near-IR photometry, IRAC data yield precise stellar
  masses and their errors ($< 0.15$ dex) for a particular choice of an
  initial mass function (IMF; e.g.,\citealp{rowan-robinson08,
    swindle11}). The primary systematic uncertainty is the unknown
  IMF; for example, changing it from the Chabrier IMF
  \citep{chabrier03} to the Salpeter IMF \citep{salpeter55} will lead
  to a shift of $\sim 0.25$ dex in stellar mass
  \citep{swindle11}. Without the IRAC data, the statistical errors in
  stellar mass would increase by a factor of two. In summary, IRAC
  observations allow us to estimate stellar masses for all of our
  observed galaxies, down to a stellar mass limit comparable to that
  reached in local clusters \citep{terlevich01,merluzzi10}.

This paper describes the survey design, key science goals, and details
of reducing the ultra deep {\Spitzer} data. We show the power of
{\surfsup} to achieve the primary goal listed above by measuring
stellar properties for a $z=9.5$ galaxy behind {\macsb}. In
\citet{surfsup2} we present details of the photometry and measurements
of the stellar masses and SFRs for $z\sim 7$ galaxies behind the
Bullet Cluster. The full analysis of all 10 clusters, which will allow
us to answer the questions described above, will be presented in
subsequent papers after the final data is taken. The paper is
structured as follows. In Section~\ref{sec:survey} we describe the
{\surfsup} program, in Section~\ref{sec:datared} we present the data
reduction steps. In Section~\ref{sec:key} we present the main science
goal of the survey.  We summarize our conclusions in
Section~\ref{sec:conclusions}. Throughout the paper we assume a
$\Lambda$CDM concordance cosmology with $\Omega_{\rm m}=0.27$,
$\Omega_{\Lambda}=0.73$, and Hubble constant $H_0=73{\rm\
  kms^{-1}\:\mbox{Mpc}^{-1}}$ \citep{komatsu11, riess11}.  Coordinates
are given for the epoch J2000.0, and magnitudes are in the AB system.

\section{Survey Design and Sample Selection}
\label{sec:survey}
The survey will use the magnification power of 10 accurately-modeled
cosmic telescopes to study galaxy populations at $z\gtrsim 1-10$ with
the main focus of studying $z\gtrsim 7$ galaxies. The clusters were selected based on a number of criteria, listed below.  
\begin{itemize}
\item The clusters need to be very efficient lenses (i.e., having
    significant areas of high magnification). This requires them to have
  large mass ($M_{500}\gtrsim 10^{15}M_{\odot}$, see
  Table~\ref{tab:clusters}) and be preferentially elliptical in
  shape. Furthermore, the critical density $\Sigma_{\rm c}$ which
  relates surface mass density $\Sigma$ to lensing convergence
  $\kappa=\Sigma/\Sigma_{\rm c}$ is larger at lower redshift,
  therefore clusters at higher redshifts are likely more efficient lenses. We select
  clusters whose areas of high magnification are well-matched to both
  the {\Spitzer} and {\it \hst/ACS} FOV. Finally, we also want to
  minimize the obscuration of background galaxies by foreground
  cluster members. Due to the smaller apparent size and brightness of
  the cluster members at higher redshifts the ideal redshifts chosen
  for this survey is around $z\sim 0.5$.\item Availability of deep
  {\hst} ACS and WFC3-IR imaging (for the very efficient lens
  RCS2-2327.4$-$0204 where the {\hst} data was not available we
  obtained the data as a part of this program).
\item Absence of bright stars in the {\Spitzer} Field-Of-View (FOV;
  we use 2MASS - \citealp{2mass} - catalog to check that no stars with K-band
    magnitude $<10$ were present near the cluster core).

\end{itemize}
Much of the work has been done in
detecting such population in {\hst}. Surveys of blank fields, in
particular HUDF, Cosmic Assembly Near-infrared Deep Extragalactic
Legacy Survey CANDELS, and the Brightest of Reionizing Galaxies Survey
BoRG (e.g., \citealp{ellis13, schenker13, bouwens12,
  oesch12,finkelstein12,grogin11,koekemoer11,mclure11, trenti11,
  yan09}) as well as cluster fields \citep{postman12, zheng12,
  coe13,zitrin12, hall12} have been undertaken. Most of the {\hst}
data needed for this project already exist, mostly through the CLASH
campaign \citep{postman12}, and through GO
observations ({\hst}-GO11591 PI Kneib, {\hst}-GO11099 PI Brada\v{c}, {\hst}-GO10846 PI
Gladders, {\hst}-GO9722 PI Ebeling). Furthermore 2 targets ({\macsos}
and \macsb) are part of the Frontier Field campaign and will be
observed in Cycle 22 with 140 orbits of {\hst} time each achieving
$\approx 28.7-29\mbox{ mag}$ in the optical (ACS) and NIR (WFC3). The
additional {\hst} observations needed for an exceptional lens
RCS2-2327.4$-$0204 have been collected as part of this program. The
cluster has been imaged in Cycle 21 with 13 prime and 13 parallel orbits
({\hst}-GO-13177, PI Brada\v{c}). The prime pointing is covered
with ACS/F814W 3 orbits, WFC3/F098M 3 orbits, WFC3/F125W 3 orbits, and
WFC3/F160W 4 orbits. We also complement the space-based observations
with deep ground-based HAWK-I $\mbox{K}_{\rm s}$ band data where available, to
even further improve constraints on stellar masses and redshifts.

The sample of galaxy clusters is presented in
Table~\ref{tab:clusters}. Many of them are merging; this is not
surprising as merging clusters have the highest projected
ellipticity and hence high lensing efficiency. In particular, high projected ellipticity in the mass
distribution generates large critical curves and large areas of high
magnifications \citep{meneghetti10}. Due to the large number of
multiply imaged systems it is usually not more difficult to model
the magnification distribution of a merging cluster compared to
the relaxed clusters. Finally, we caution that due to the merging
nature of many of the clusters the masses quoted in
Table~\ref{tab:clusters} might be overestimated; this, however, does
not influence the selection as we modeled the magnification distribution separately with the main goal to select the
clusters with the highest lensing efficiency within a WFC3-IR FOV.

\setlength{\tabcolsep}{3pt}
\begin{deluxetable*}{p{4pt}lllllllllll}
\tablecolumns{12}
\tablewidth{0pt}
\tablecaption{Target list. }
\tablehead{\colhead{} & \colhead{Target Name} & \colhead{RA} &
  \colhead{Dec}   & \colhead{z} & \colhead{$kT$}&\colhead{$r_{500}$} &\colhead{$r_{500}$}&\colhead{$M_{500}$} &$M$& \colhead{Total}&\colhead{Archive}\\
\colhead{} & \colhead{} &\colhead{} &\colhead{} &\colhead{} &\colhead{[keV]} &\colhead{Mpc}&\colhead{arcmin} &\colhead{$[10^{15}M_{\odot}]$} &\colhead{\tablenotemark{[*]}}&\colhead{exp.} &\colhead{}} 
\cline{1-10}
\startdata 
1 &MACSJ0454.1$-$0300&04:54:10.90&$-$03:01:07.0      &0.54 & $7.5 \pm 1.0^{\mbox{(a)}}$         &   $1.31 \pm 0.06$ & $3.54 \pm 0.16 $&$1.15 \pm 0.15^{\mbox{(i)}}$   &2&$114\mbox{ks}$ & $24 \mbox{ks}$\tablenotemark{1,2,3}\\
2 &Bullet Cluster   &06:58:27.40&$-$55:56:47.0      & 0.30 & $11.70 \pm 0.22^{\mbox{(b)}}$     &   $1.81 \pm 0.07$ &  $7.02 \pm 0.27 $&$2.28 \pm 0.28^{\mbox{(i)}}$   &4& $111\mbox{ks}$ & $22 \mbox{ks}$\tablenotemark{1,4,5}\\      
3 &MACSJ0717.5$+$3745&07:17:33.80&$+$37:45:20.0   & 0.55 &$12.5 \pm 0.70^{\mbox{(a)}}$         &   $1.69 \pm 0.06$ &  $4.53 \pm 0.16 $&$2.49 \pm  0.27^{\mbox{(i)}}$  &4&$108\mbox{ks} $  & $18 \mbox{ks}$\tablenotemark{1}\\
4 &MACSJ0744.8$+$3927&07:44:51.80&$+$39:27:33.0   &0.70 &$8.9 \pm 0.80^{\mbox{(a)}}$           &   $1.26 \pm 0.06$ &  $3.02 \pm 0.14 $&$1.25 \pm 0.16^{\mbox{(i)}}$   &1/2&$108\mbox{ks}$ & $18 \mbox{ks}$\tablenotemark{1}\\
5 &MACSJ1149.5$+$2223&11:49:34.30&$+$22:23:42.0 & 0.54 & $8.7 \pm 0.90^{\mbox{(a)}}$          &   $1.53 \pm 0.08$ &   $4.14 \pm 0.22 $&$1.87 \pm 0.30^{\mbox{(i)}}$   &4& $108\mbox{ks}$ & $18 \mbox{ks}$\tablenotemark{1}\\
6 &RXJ1347$-$1145&13:47:32.00&$-$11:45:42.0          & 0.59 & $10.75 \pm 0.83^{\mbox{(b)}}$   &    $1.67 \pm 0.08$&   $4.32 \pm 0.21 $&$2.17 \pm 0.30 ^{\mbox{(i)}}$  &2& $113\mbox{ks}$ & $23\mbox{ks}$\tablenotemark{1,2}\\
7 &MACSJ1423.8$+$2404&14:23:48.30&$+$24:04:47.0 &  0.54 &$7.1 \pm 0.65^{\mbox{(c)}}$          &   $1.09 \pm  0.05$ &  $2.95 \pm 0.14 $&$0.66 \pm  0.09^{\mbox{(i)}}$  &1& $108\mbox{ks}$ & $18\mbox{ks}$\tablenotemark{1}\\   
8 &MACSJ2129.4$-$0741&21:29:26.21&$-$07:41:26.2    & 0.59  &$9.0 \pm 1.20^{\mbox{(a)}}$        &   $1.25 \pm 0.06$ &  $3.24 \pm 0.16 $&$1.06 \pm 0.14^{\mbox{(i)}}$   &3& $108\mbox{ks}$ &$18 \mbox{ks}$\tablenotemark{1}\\
9 &MACSJ2214.9$-$1359&22:14:57.41&$-$14:00:10.8  & 0.50 &$8.8 \pm  0.7^{\mbox{(a)}}$          &   $1.39 \pm 0.08$ &   $3.92 \pm 0.23 $&$1.32 \pm 0.23^{\mbox{(i)}}$   &2& $108\mbox{ks}$& $18 \mbox{ks}$\tablenotemark{1}\\
10 &RCS2-2327.4$-$0204&23:27:28.20&$-$02:04:25.0 & 0.70 & $9.5^{+1.8\; \mbox{(d)}}_{-1.2}$      & $1.16^{+0.11}_{-0.08}$ &$2.78^{+0.26}_{-0.19}$& $1.23^{+0.16}_{-0.15}$$\phantom{.}^{\mbox{(ii)}}$&2&$108\mbox{ks}$ & $18 \mbox{ks}$\tablenotemark{1}
\enddata
\tablenotetext{1}{P60034: PI Egami: ``The IRAC Lensing Survey: Achieving JWST depth with {\Spitzer}''}
\tablenotetext{2}{P00083: PI Rieke: ``Use of Massive Clusters
  as Cosmological Lenses/Evolution of Galaxies and Lensing in
  Clusters''}
\tablenotetext{3}{P50610: PI Yun: ``Charting Cluster Mass Build-up
  using Luminous IR Galaxies''}
\tablenotetext{4}{P03550: PI Jones: ``Star Formation and Galaxy Evolution During a Supersonic Cluster Merger''}
\tablenotetext{5}{P40593: PI Gonzalez : ``Quenched Star Formation in
  the Bullet Cluster''}
\tablenotetext{*}{As in \protect{\citet{mann12}} morphology $M$ is assessed visually based on the appearance of the X-ray contours and the goodness of the optical/X-ray alignment. The assigned codes are from 1 - apparently relaxed to 4 - extremely disturbed. References: \citet{allen08,mann12,vonderlinden12}}
\tablecomments{Total and archive exposures are given per channel.  X-ray temperature and redshift references (a) \protect{\citet{ebeling07}}  (b) \protect{\citet{mantz10}} (c) \protect\citet{postman12} (d) Gladders et al. in prep. \\$M_{500}$, and $r_{500}$ are derived from X-ray data; references (i) \protect{\citet{mantz10}}, (ii) Gladders et al. in prep.}
\label{tab:clusters}
\end{deluxetable*}

\section{Spitzer Data Reduction and Properties}
\label{sec:datared}

The observations of all clusters were taken in 4 scheduling
blocks, two blocks (separated by $\sim 10\deg$ in the roll angle) were
followed by two more, separated by $\sim 180\deg$ from the previous
two to ensure coverage in both channels in the flanking fields.  Two
pointings (one pointing per band) were sufficient to cover the entire
region of high magnification ($\mu > 2$).

Our basic data processing
begins with the corrected-basic calibrated data (cBCD).  These data
include a few IRAC artifact-correction procedures.  However, visual
inspection of preliminary mosaics illustrates that additional
mitigation measures are required.  Therefore we applied the
warm-mission column pulldown (\texttt{bandcor\_warm.c} by M.~Ashby)
and an automuxstripe correction contributed software
(\texttt{automuxstripe.pro} by
J.~Surace)\footnote{\url{http://irsa.ipac.caltech.edu/data/SPITZER/docs/dataanalysistools/tools/contributed/irac/}}
to the individual cBCDs from both channels.  These steps produce
noticeably improved mosaics, particularly near the very bright stars.
While there are no very bright stars near the cluster core, there are
some in the flanking fields.

The process of creating the mosaic images closely follows the IRAC
Cookbook\footnote{\url{http://irsa.ipac.caltech.edu/data/SPITZER/docs/dataanalysistools/cookbook/}}
for the COSMOS medium-deep data; here we describe a few noteworthy
exceptions.  Like in the Cookbook, all processing from here on is
performed with the MOsaicker and Point source EXtractor (\mopex)
command-line
tools\footnote{\url{http://irsa.ipac.caltech.edu/data/SPITZER/docs/dataanalysistools/tools/mopex/}}.
The {\it overlap correct} is applied to all cBCD frames to bring their sky
backgrounds to agreement across the final mosaic \citep{makovoz05}.
For this correction, we use the {\tt DRIZZLE} option for interpolation
with $\mbox{pixfrac}=1$ to fully cover the output pixels.
Although this interpolating procedure is considerably slower than
others (e.g., spline or bicubic), it produces mosaics with cleaner sky
backgrounds.  The overlap correction generates temporary files which
are used in the next stage of processing.

The two flanking fields are adjacent and aligned with the primary
IRAC pointing. Their positions were determined by spacecraft
visibility.  They typically have only half the number of frames of the
primary field. Therefore, we generate two different mosaics per
channel. We have typically $\sim\!2000$~individual frames for the
central region, so we use the {\tt DRIZZLE} algorithm with
$\mbox{pixfrac}\!=\!0.01$ to interpolate the overlap-corrected cBCDs
onto the output mosaic.  This mosaic results in severe holes and noise
along the edges where many fewer frames are available --- including the interior edges between the primary
and flanking fields.  We therefore produce a
second mosaic with $\mbox{pixfrac}\!=\!0.85$, to provide clean images
in all connected regions. The former is better for objects close to
the cluster core, while the latter is useful when imaging in the
flanking fields is required. Both mosaics of the Bullet Cluster are
available to the public (see Sec.~\ref{sec:datarel}). We will do the
same for all reduced data for the remaining 9 targets in the future.

\begin{figure}[hb]
\begin{center}
\includegraphics[width=0.5\textwidth]{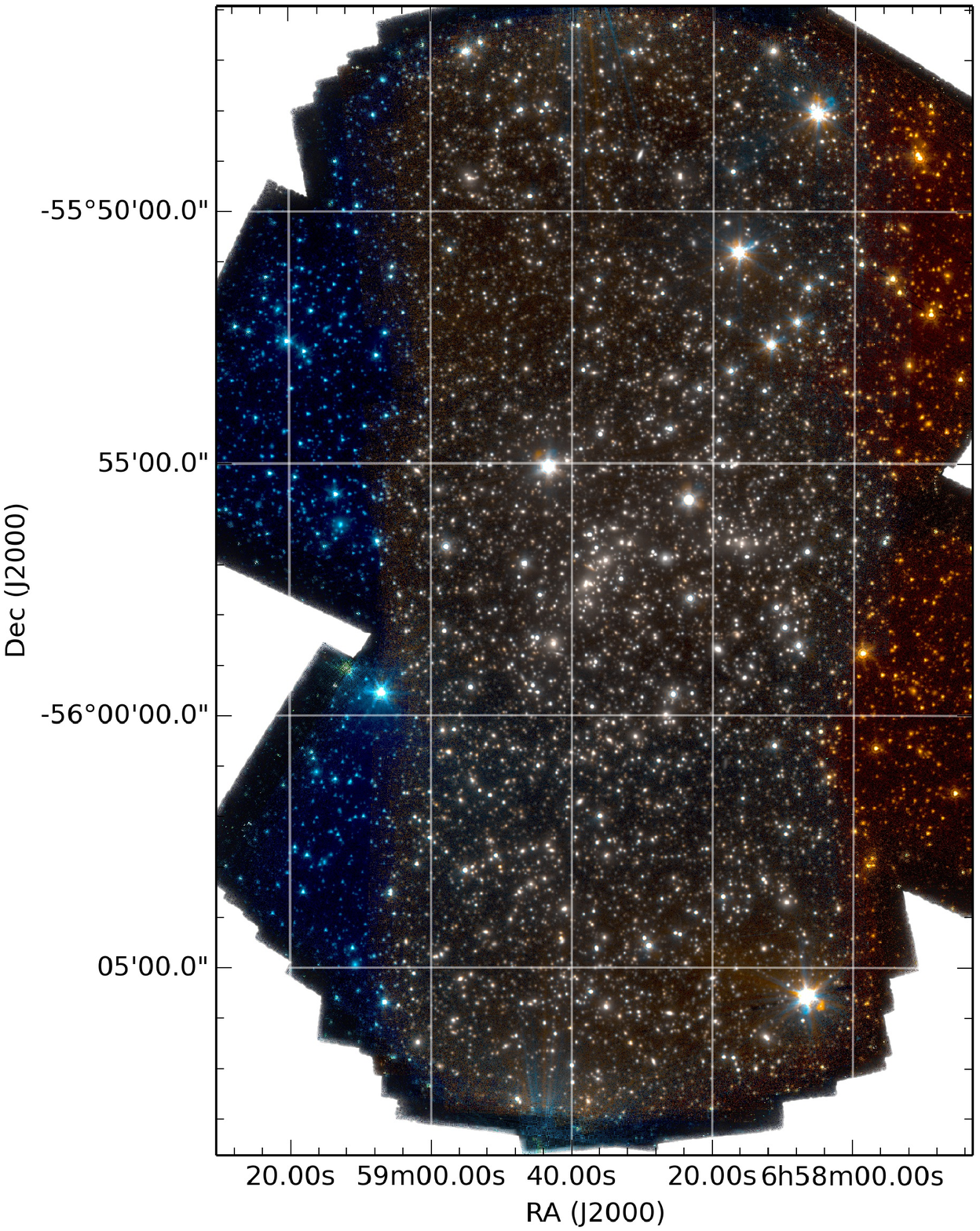}
\end{center}
\caption{False color map of the Bullet Cluster data
  using {\Spitzer} ${\chtwo}$ (red), ${\chtwo}+{\chone}$ (green), and
  ${\chone}$ (blue) data as the RGB channels. Areas where only
  ${\chone}$ (${\chtwo}$) data are available (from earlier programs) are clearly visible in blue
  (orange). The figure was produced using {\tt STIFF}{\protect\footnote{\url{http://www.astromatic.net/software/stiff}}} and {\tt APLpy}
  package{\protect\footnote{\url{http://aplpy.github.com}}}. This map uses FITS
  images created with $\mbox{pixfrac}\!=\!0.01$ (see text).
}
\label{fig:map}
\end{figure}

\begin{figure*}[ht]
\begin{center}
\includegraphics[width=0.8\textwidth]{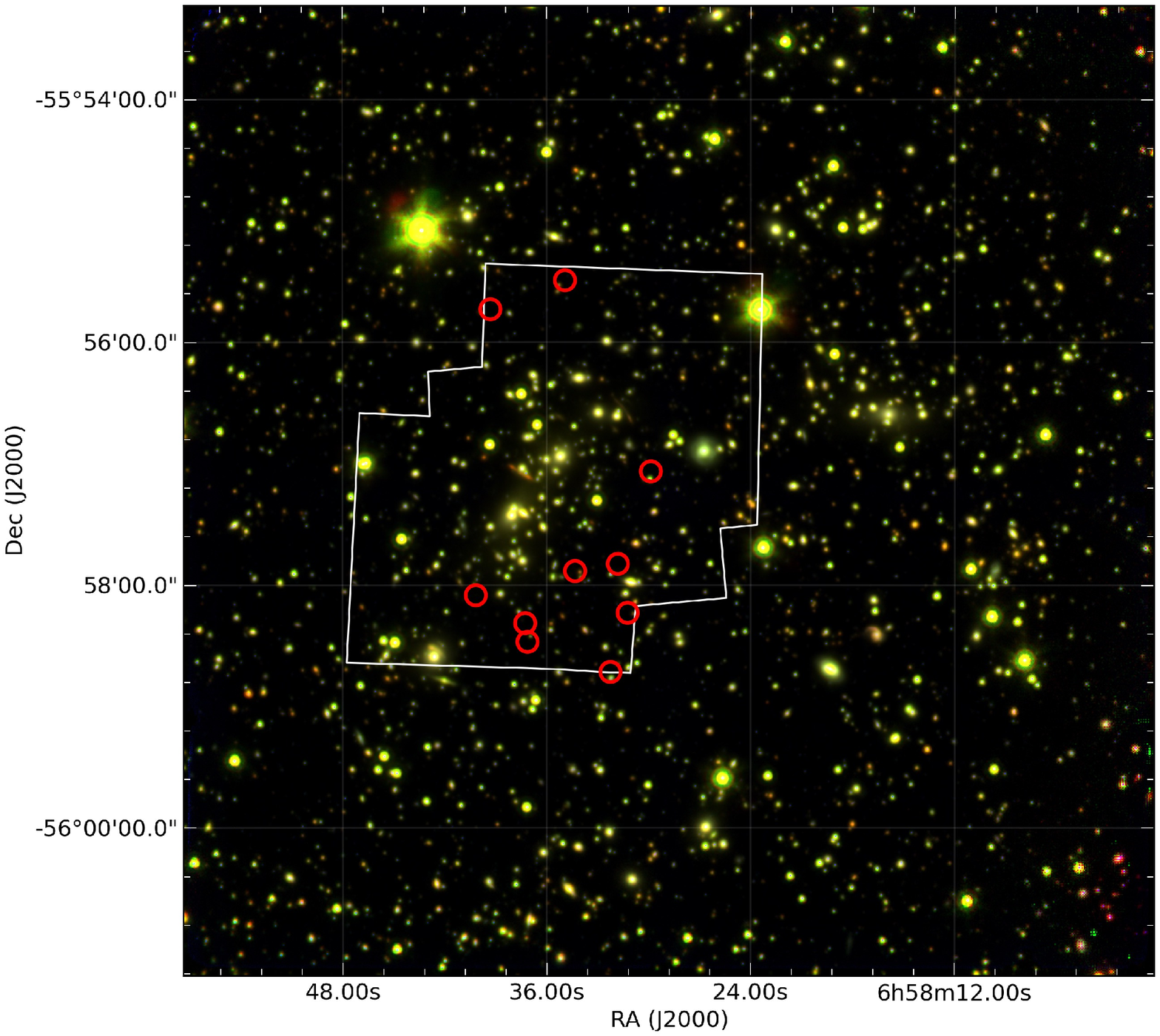}
\end{center}
\caption{A zoomed-in color map using {\Spitzer} ${\chtwo}$, ${\chone}$ and HAWK-I $\mbox{K}_{\rm s}$ band data as RGB channels. Overlaid is the  {\hst}   F160W  footprint (white polygon),  and
\citet{hall12} $z\sim 7$ candidates (red circles). The figure was produced
  following the algorithm from \citet{lupton04} and using {\tt APLpy}
  package.}
\label{fig:map2}
\end{figure*}

We also use the available archival data to produce the final
mosaics; for example, the \surfsup\ {\Spitzer}/IRAC data for the Bullet Cluster
({\bullet}, \citealp{tucker95}) was complemented with
existing data from two programs: 3550 (PI:~C.~Jones, cryo-mission) and
60034 (PI:~E.~Egami, warm-mission).
 For the Bullet Cluster in total there are $\sim\nframes$ individual frames (per channel),
with each having a nominal frametime of 100~s. The final mosaics have
a pixel scale of $0\farcs60$~pix$^{-1}$ (an integer multiple of the
{\hst} pixel scale) and have a position angle of
$\texttt{CROTA}\!=\!0\fdg$  By comparing the {\Spitzer} and {\hst}
positions of bright objects we correct for any residual shifts in the
relative astrometry (for the Bullet Cluster
$[\Delta\alpha,\Delta\delta]\!=\![+0\farcs18,-0\farcs12]$), and we
subtract it from the \texttt{CRVAL} keywords of the {\Spitzer}
images. In \fig{fig:map}, we show the false color image using both
channels, \fig{fig:map2} shows a zoomed-in color map of the Bullet Cluster using {\Spitzer}
and HAWK-I $\mbox{K}_{\rm s}$ band \citep{clement12} with the {\hst}
F160W footprint, and \citet{hall12} $z\sim 7$ candidates overlaid.

\subsection{Depth and Sky Background \label{sec:rms}} 

\begin{figure*}[h]
\begin{center}
\begin{tabular}{cc}
\includegraphics[width=0.5\textwidth]{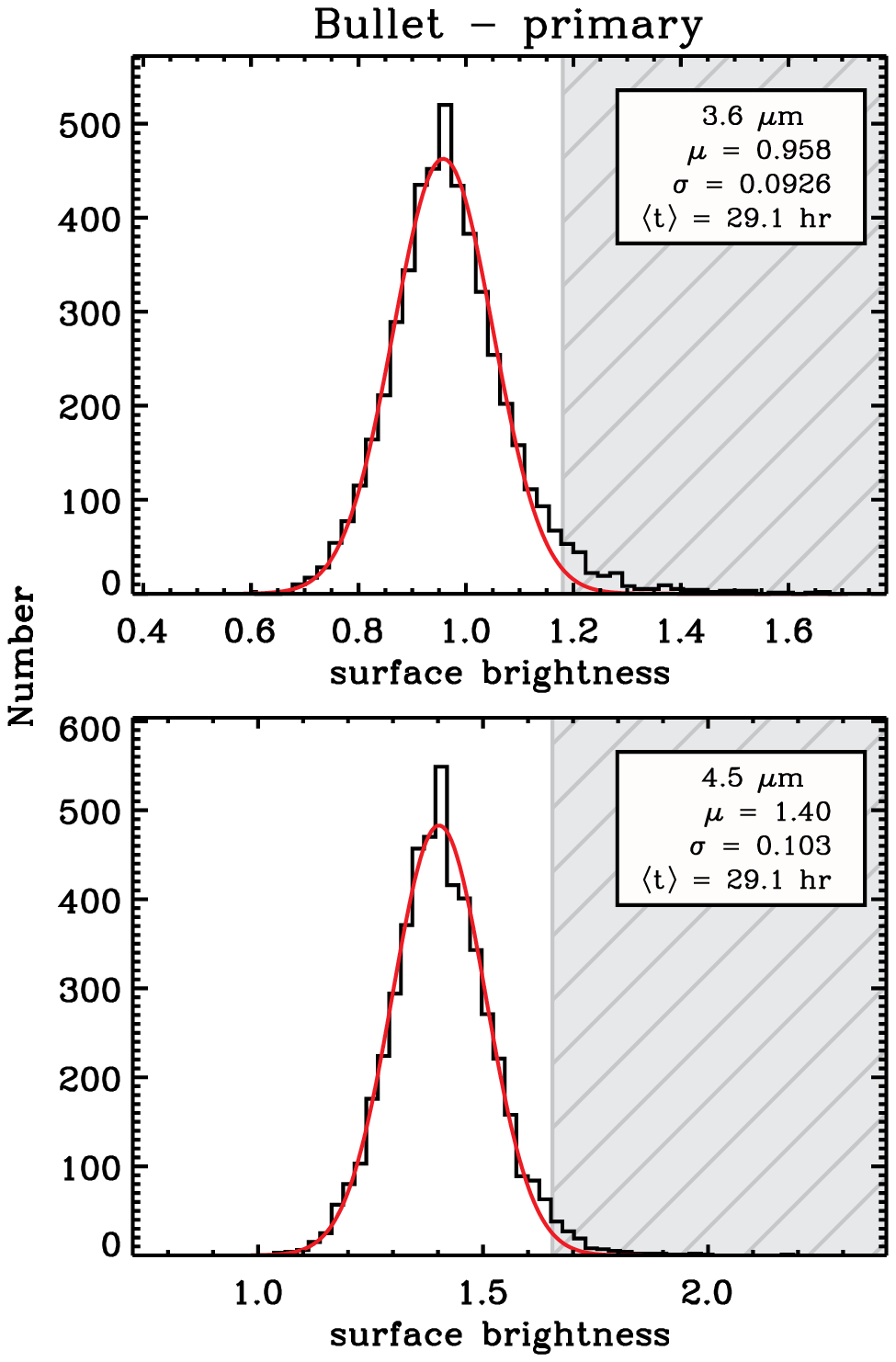} & \includegraphics[width=0.5\textwidth]{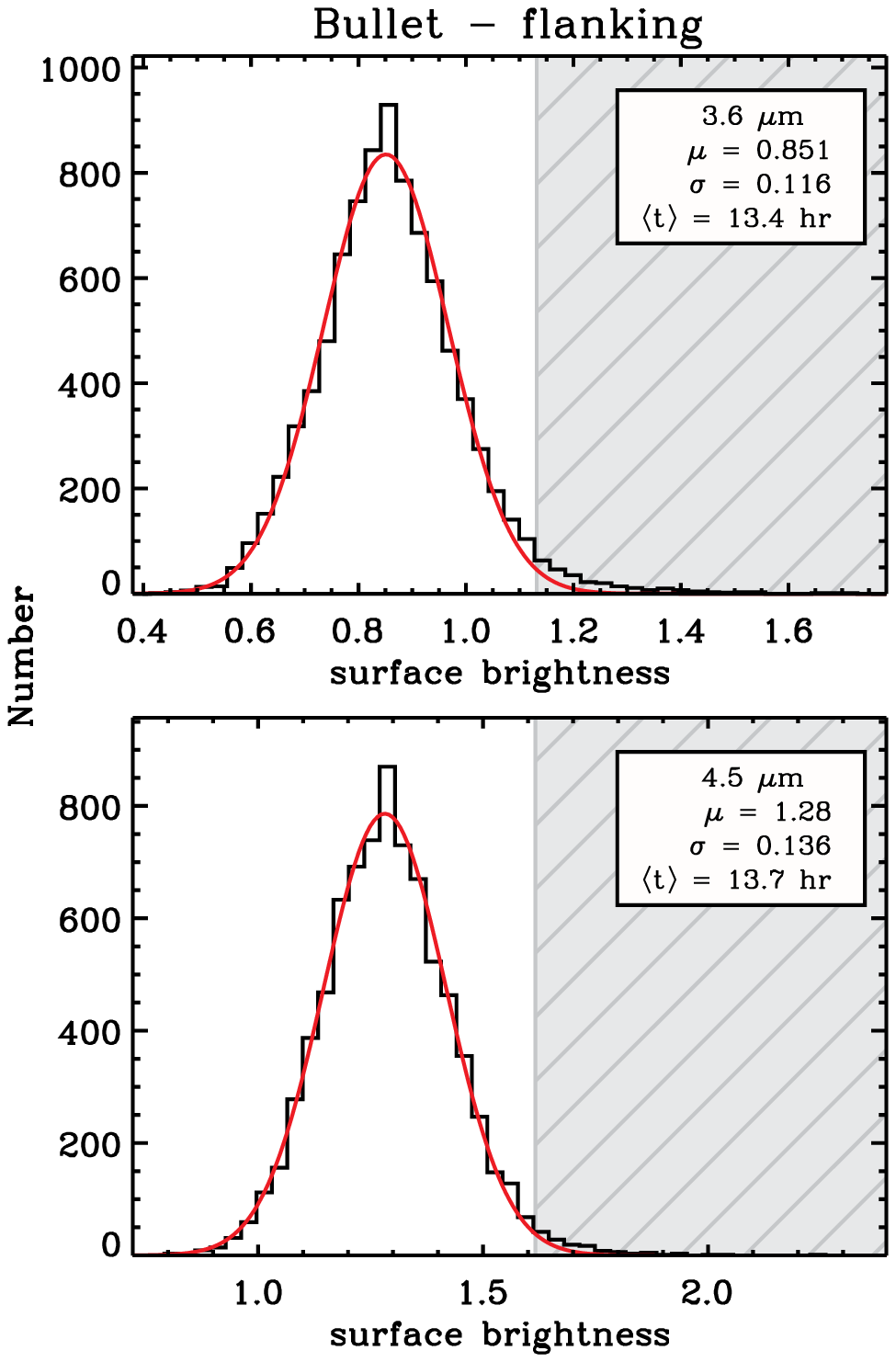}
\end{tabular}
\end{center}
\caption{Distribution of sky surface brightness for ${\chone}$ ({\bf
    top}) and ${\chtwo}$ ({\bf bottom}) data for the Bullet
  cluster. In the {\bf left} column are measurements for the primary,
  and the {\bf right} for the flanking field.  We have measured the
  average ($\mu$) and RMS ($\sigma$) of the sky in non-overlapping
  boxes obviously free of any objects. The histograms are centered on
  the average local sky value $\mu$ for all realizations (see
  Sect.~\ref{sec:rms}).  It is clear that there is still a very
  low-level contamination from faint sources, from the positively
  skewed tail.  To estimate the RMS, we fit a Gaussian distribution
  omitting the contaminated region shown in grey.}
\label{fig:rms}
\end{figure*}

We measure the sky statistics from $>\!50$ non-overlapping boxes
placed in regions of roughly equal exposure time.  These boxes
typically contain $\sim\!100$~pixels and are chosen to be devoid of
any objects (or object wings) or significant intra cluster light
contribution (the latter is seen as increased background level close
to the cluster center).  We compute the average sky surface brightness
with 4-$\sigma$ outlier rejection separately for each box.  We combine
the sky-subtracted boxes into a single histogram and add back the
global average of the sky surface brightnesses (note that the global
background has {\it not} been subtracted from the images).  In
Figure~\ref{fig:rms} we show the distribution of sky surface
brightnesses for both IRAC bands (rows) and primary/flanking fields
(left/right columns, respectively).  The red curve represents a
Gaussian fit to the observed sky surface brightness distribution, and
the hatched region indicate a positive tail omitted from this fitting.
This positive tail is likely due to very faint wings or marginally
detected objects in the sky boxes.  We give the root-mean squared
(RMS) of these Gaussian models for both IRAC bands and
primary/flanking fields for the eight clusters with data available at
the time of submission (see Table~\ref{tab:improp}).

One potential drawback to using crowded fields is that the flux from
 additional unresolved sources, extended low surface brightness
  sources, and the wings of bright objects can cause higher RMS values
  than in uncrowded fields. However, our RMS values agree well with
  the exposure time calculator
  (ETC)\footnote{\url{http://ssc.spitzer.caltech.edu/warmmission/propkit/pet/senspet/}}
  predictions. ETC estimates of the sensitivity for the Bullet Cluster indicate that we should be marginally less sensitive (by $\sim 20\%$)
  in $\chone$ and equally sensitive in $\chtwo$ compared to our measurements.
Furthermore, we have measured RMS using GOODS v0.3 mosaics (data taken
during cryogenic mission) and it is in agreement with our flanking
field values (which have similar exposure times). We have furthermore
reduced the full-depth ($40\mbox{hr}$) data for GOODS and using simple
scaling (rescaling RMS with ${t_{\rm exp}}^{0.5}$) the predicted RMS
is comparable to {\surfsup} (it is lower in $\chone$ and higher in
$\chtwo$, see Table~\ref{tab:improp}). The differences can be caused
by higher background, improper rescaling to match exposure time,
higher contamination from local sources, and/or instrument
degradation.

The massive foreground clusters could in principle degrade the
{\surfsup} survey depth via source confusion effects. We therefore
measured the sensitivity limits of the {\surfsup} mosaics by placing
artificial sources in the vicinities of the objects of interest and
measuring their fluxes in addition to calculating the RMS sky values
in their vicinities. We also compute the RMS in a $3\arcsec$ (radius)
aperture on the sky after “cleaning” the foreground objects. Using RMS
values, for the Bullet Cluster we achieve 3-$\sigma$ limiting
magnitudes of $\sim 26.6$ in ${\chone}$ and $\sim 26.2$ in ${\chtwo}$
(the exact value is dependent on the location of the source and is
similar for the two methods). Finally, as noted by \citet{ashby13},
the common practice of basing photometric uncertainties on such noise
estimates is problematic, because of a possible residual flux from
unresolved sources. By measuring background levels across the mosaic, we estimate the uncertainty
due to unresolved sources to be of the order
$0.2-0.5\mbox{mag}$. We conclude that the degradation is not
significant and is more than compensated by the magnification of the
cluster.  More discussion of the photometry is presented by
\citet{surfsup2}.

\begin{deluxetable*}{lcccccc}
\tablecolumns{7}
\tablewidth{\textwidth}
\tablecaption{Image properties }
\tablehead{\colhead{Target Name} & \colhead{RMS ${\chone}$} & \colhead{RMS ${\chtwo}$} & \colhead{RMS ${\chone}$} & \colhead{RMS ${\chtwo}$} & \colhead{PSF FWHM $\chone$} &  \colhead{PSF FWHM $\chtwo$} \\ \colhead{} & \colhead{$[10^{-3} \mbox{MJy/sr}]$} & \colhead{$[10^{-3} \mbox{MJy/sr}]$} & \colhead{$[10^{-3} \mbox{MJy/sr}]$} & \colhead{$[10^{-3} \mbox{MJy/sr}]$} & \colhead{[pixel]} &  \colhead{[pixel]} \\ \colhead{} & \colhead{Primary Field} & \colhead{Primary Field} & \colhead{Flanking Field} & \colhead{Flanking Field}&\colhead{$1 \mbox{pix}=0.60"$} & \colhead{$1 \mbox{pix}=0.60"$}}
\cline{1-7}
\startdata 
Bullet Cluster      & 0.931 & 1.04 & 1.17 & 1.36 &  2.82 & 2.72\\
MACSJ0454.1$-$0300   & 0.992 & 1.19 & 1.30 & 1.59 &  2.76 & 2.79\\
MACSJ0717.5$+$3745   & 0.957 & 1.14 & 1.20 & 1.55 &  2.83 & 2.83\\
MACSJ0744.8$+$3927   & 0.904 & 1.09 & 1.12 & 1.49 &  2.91 & 2.87\\
MACSJ1149.5$+$2223   & 0.905 & 1.10 & 1.08 & 1.44 &  2.76 & 2.87\\
MACSJ1423.8$+$2404   & 0.757 & 0.98 & 1.06 & 1.29 &  2.94 & 2.91\\  
MACSJ2129.4$-$0741   & 0.840 & 1.05 & 1.10 & 1.45 &  2.83 & 2.76\\
MACSJ2214.9$-$1359   & 0.890 & 1.10 & 1.08 & 1.46 &  2.83 & 2.79\\
GOODS   & 0.708\tablenotemark{(1)} &1.44\tablenotemark{(1)}  & 1.29 & 1.64 & 2.49 & 2.46 
\enddata
\tablenotetext{1}{The GOODS data used for primary field comparison has
  a depth of 40hr, we rescaled the RMS to 29hr depth assuming it
  scales as $\propto \sqrt{t_{\rm exp}}$. For flanking fields we use
  GOODS v0.3 public data which have comparable depths.}
\label{tab:improp}
\end{deluxetable*}

\subsection{Point-Spread Function (PSF)\label{sec:psf}}
For the combined {\hst} and IRAC photometry, we generate empirical
IRAC PSFs by stacking point-sources in the field. We begin with
SExtractor \citep{sextractor} tuned to highly deblend these confused
IRAC images, specifically $\texttt{DEBLEND\_MINCONT}\!=\!10^{-5}$.
From these catalogs, we identify stars based on the correlation of
\texttt{FLUX\_RADIUS} and \texttt{MAG\_AUTO} (e.g., \citealp{ryan11},
Figure~2) requiring axis ratio of $b/a\!\geq\!0.9$. We refine the
centroids from SExtractor by fitting a 2-D Gaussian and align each
point-source with {\tt sinc} interpolation.  We mask neighboring
objects using the segmentation maps from SExtractor grown by 2~pixels
in radius.  We estimate the flux of each point-source after sky
subtraction of a sigma-clipped mean and using a circular aperture of 4
pixel radius.  At various stages, we reject point sources with bad
centroid refinement, too many masked neighbors, or suspect sky levels.
Before median-combining the shifted point sources, we normalize their
total flux to unity.  We median combine the valid sources,
and perform a second sky subtraction and flux renormalization.  We
estimate the full-width at half-maximum (FWHM) by fitting a Gaussian
to the 1-D profile. They are listed in Table~\ref{tab:improp} and are
consistent with \citet{gordon08}. We confirm a subset of point sources
that are located in both the {\HST} and {\Spitzer} data ($\sim 5$ for
a typical cluster).  Our empirical PSF FWHM values are also in
agreement with the values reported in the IRAC handbook
($1.66\arcsec=2.77 \mbox{ pixel}$ and $1.72\arcsec = 2.87 \mbox{
  pixel}$ for $\chone$ and $\chtwo$ respectively). We are releasing
the FITS images of the stacked PSF as discussed below.

\subsection{Public Data Release}\label{sec:datarel}

This program will be of use for the broader community for the study of
distant, magnified sources and IR properties of lower-redshift
galaxies and galaxy cluster members. We have waived any proprietary
rights for this program. Furthermore, we are making high-level science
products available following publication of the full data-set. We are releasing  mosaics with two
different values of $\mbox{pixfrac}\!=\!0.01$ and $0.85$ for the first
cluster.  As discussed above, the smaller pixfrac is better for objects close
to the cluster core, while the larger one is useful when imaging in the
flanking fields is required. We are also releasing the
empirical PSF FITS files, because these are needed for joint optical and
Spitzer photometry.  The data for the Bullet Cluster (and the remaining clusters in
the near future) can be found
on-line\footnote{\url{http://www.physics.ucdavis.edu/~marusa/SurfsUp.html}}. We
plan to release similar products for all the clusters in the sample.

\section{Star formation at $z\gtrsim 7$}\label{sec:key}
\begin{figure}[ht]
\includegraphics[width=0.55\textwidth]{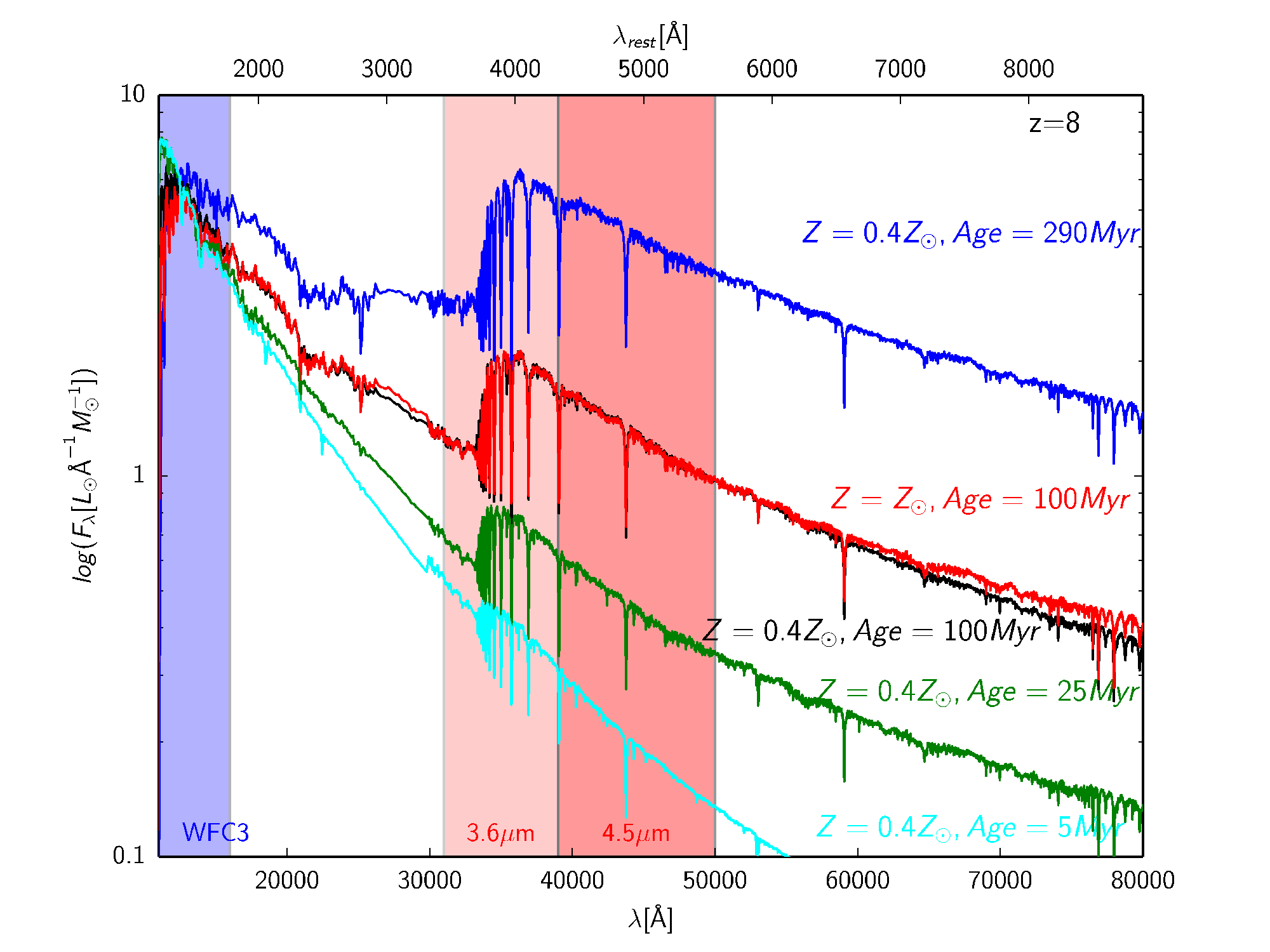} 
  \caption{Five different spectra for starburst galaxies
    (from \citealp{bruzual03}) redshifted to $z=8$. The blue curve represents a stellar population at $t=290\mbox{ Myr}$ after the burst, the red and black 
    curve are for $t=100\mbox{ Myr}$, the green one for $t=25\mbox{ Myr}$ and
    the cyan for $t=5\mbox{ Myr}$. All curves are calculated for
    a metallicity $Z=0.4Z_{\odot}$, except for the black curve
    where we use $Z=Z_{\odot}$ and $t=100\mbox{ Myr}$ (to show the
    effect of metallicity degeneracy with age which is small). Whereas
    all these galaxies would have similar colors in the {\hst}/WFC3 bands
    (blue shaded region, similar spectral slopes within photometric
    uncertainties), the different ages can be easily distinguished once
    ${\chone}$ and ${\chtwo}$ {\Spitzer} imaging is added
    (red shaded region), as their $H_{\rm{160W}}-[{\chone}]$ and
    $H_{\rm{160W}}-[{\chtwo}]$ colors are very different and hence
    their stellar masses and ages can be determined reliably.}
\label{fig:spect}
\end{figure}

As mentioned above, the key science goal of {\surfsup} is the study of
the properties (star formation rates and stellar masses) of a
representative sample of galaxies. Fig.~\ref{fig:spect} shows five
model starburst galaxies with different stellar ages and
metallicities. While these galaxies would not be detected in the
optical and have similar colors in WFC3/IR bands, they show large
differences in the ${H_{\rm{160W}}}-[{\chone}]$ and
${H_{\rm{160W}}}-[{\chtwo}]$ colors. The redshift is mostly determined
by the detection in the WFC3-IR and non detection in the bluer bands;
{\Spitzer} data is crucial to determine stellar ages and masses.  In
\citet{surfsup2} we present the detailed {\Spitzer} photometry and
stellar properties for $z$-band dropouts behind the Bullet Cluster
from \citet{hall12}. Here we describe a detection and measurement of
the stellar properties of the $z = 9.5$ galaxy behind
MACSJ1149.5$+$2223 (MACS1149-JD) from \citet{zheng12}.
\subsection{Stellar properties of MACS1149-JD}\label{sec:1149}

\begin{figure}[ht]
\begin{center}
\begin{minipage}{0.4\textwidth}
\begin{tabular}{ccc}
\multirow{2}{*}[1cm]{\includegraphics[width=0.33\textwidth]{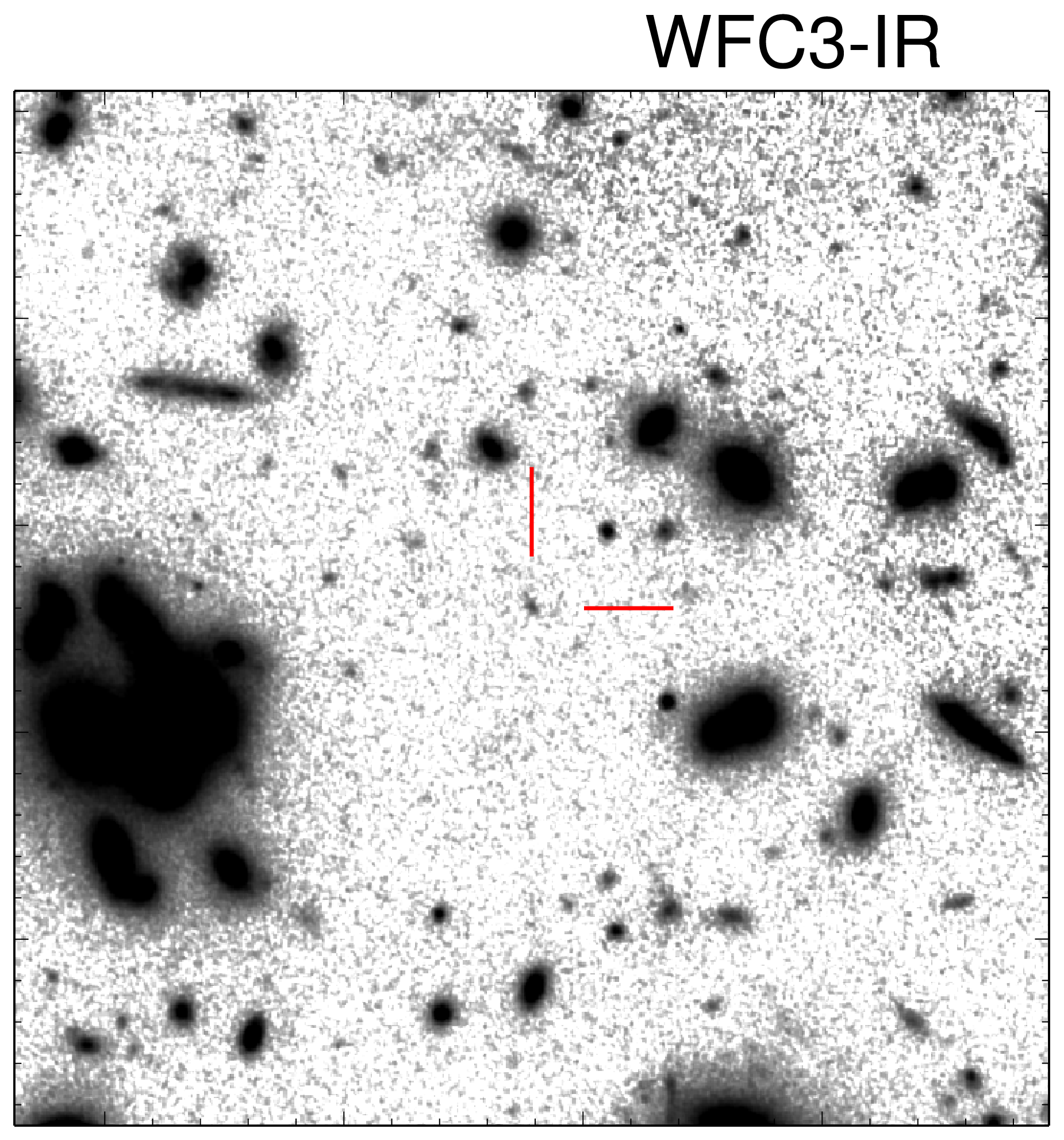}}& \includegraphics[width=0.33\textwidth]{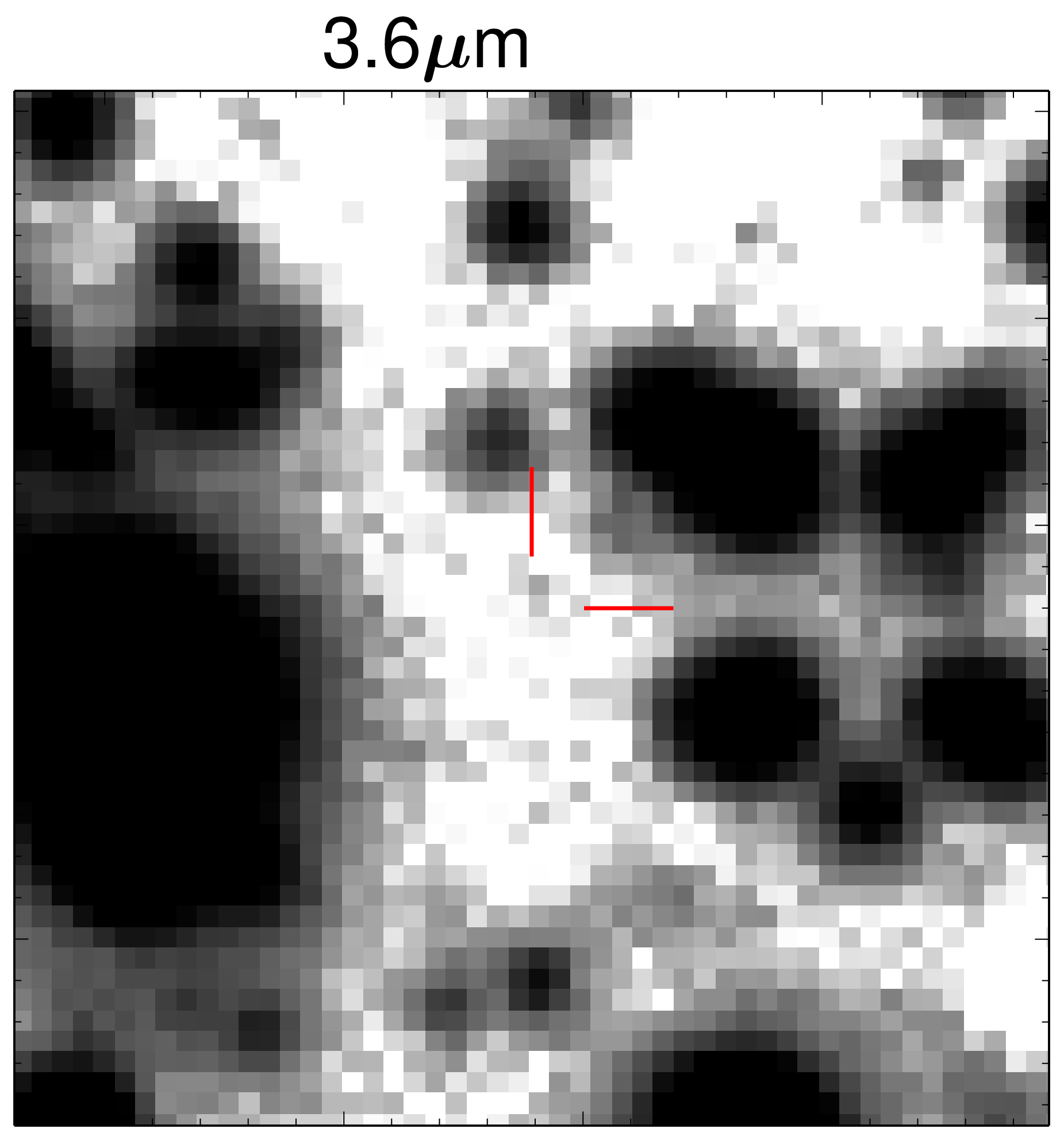}& \includegraphics[width=0.33\textwidth]{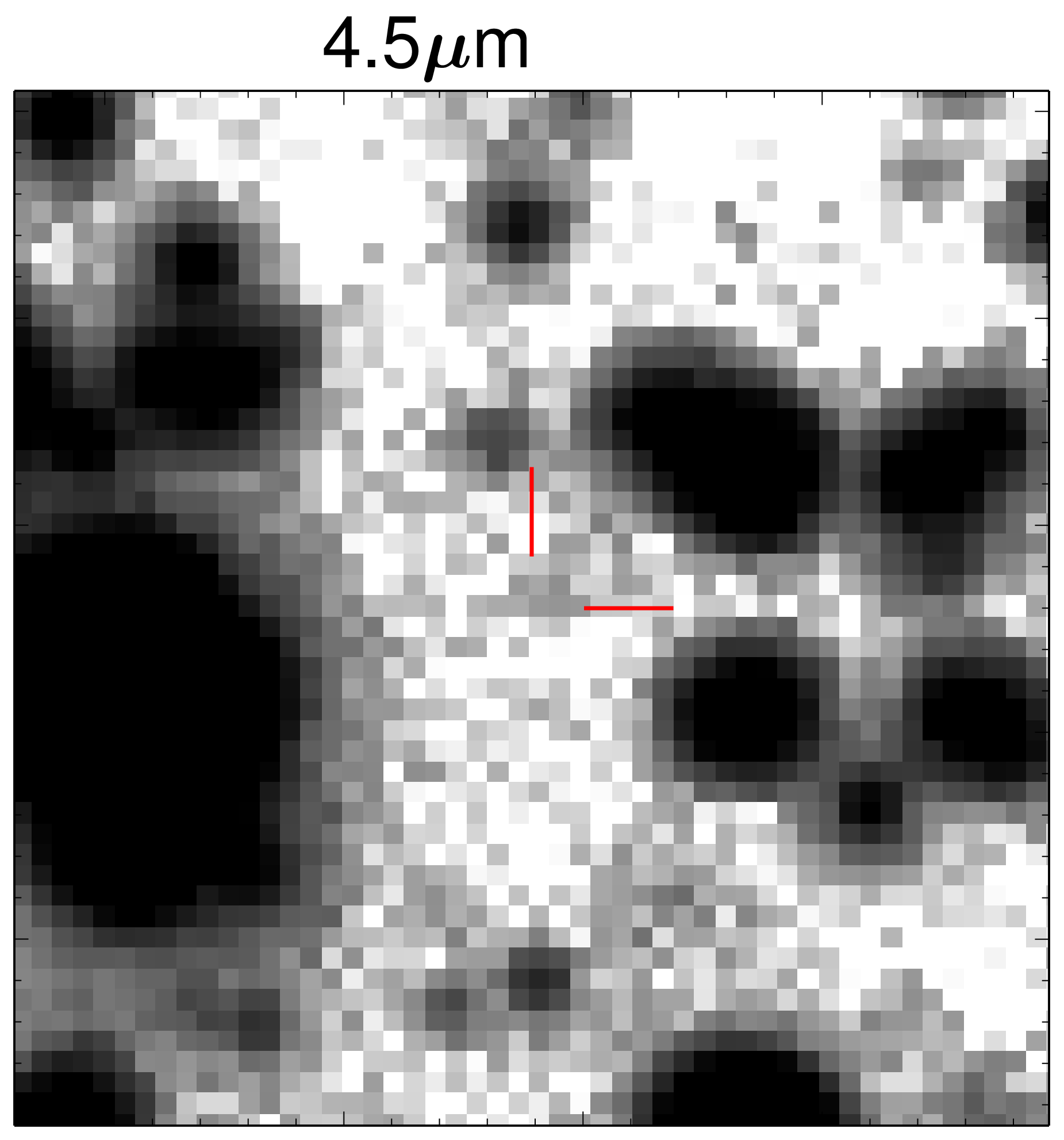}\\
& \includegraphics[width=0.33\textwidth]{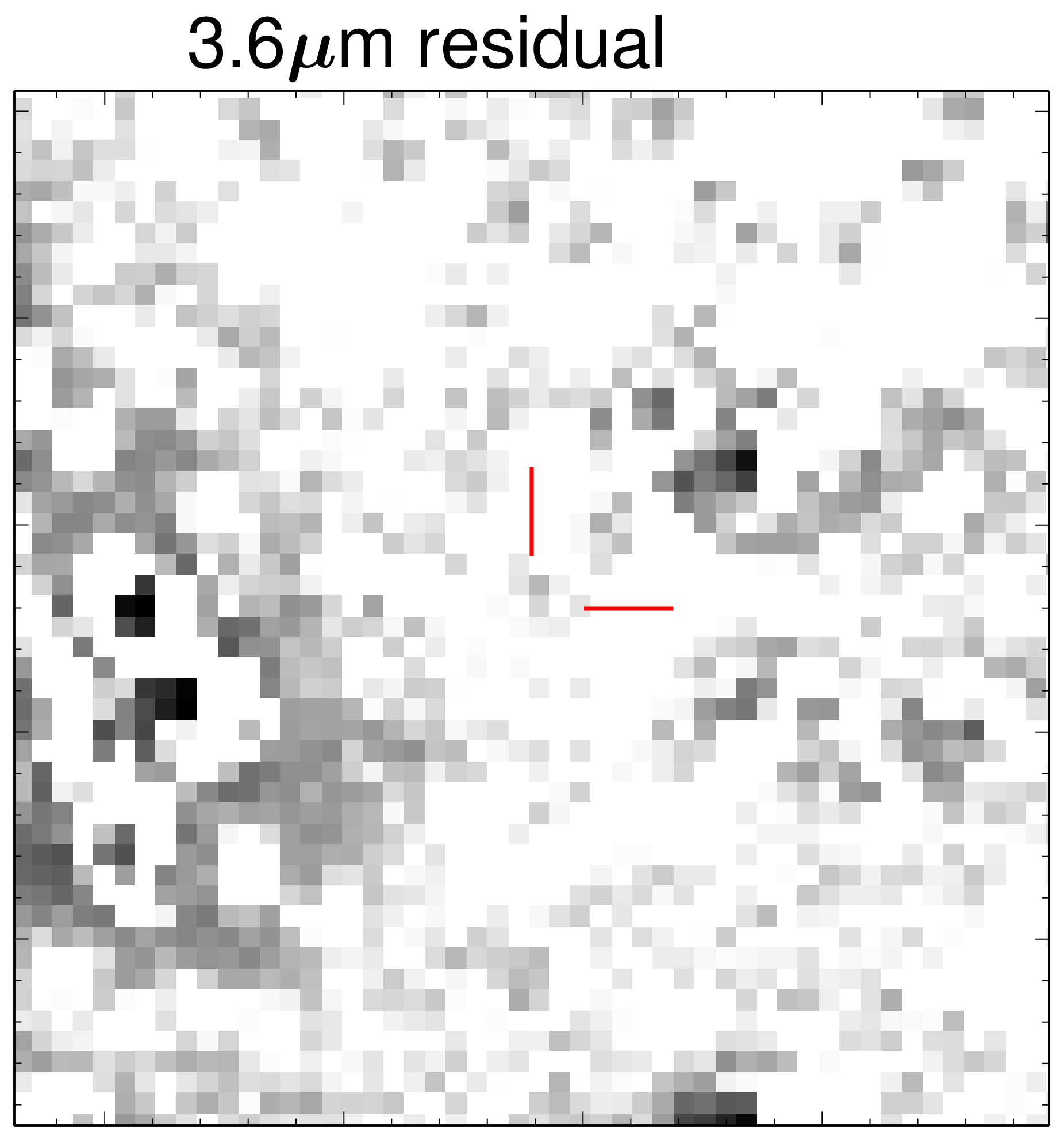} & \includegraphics[width=0.33\textwidth]{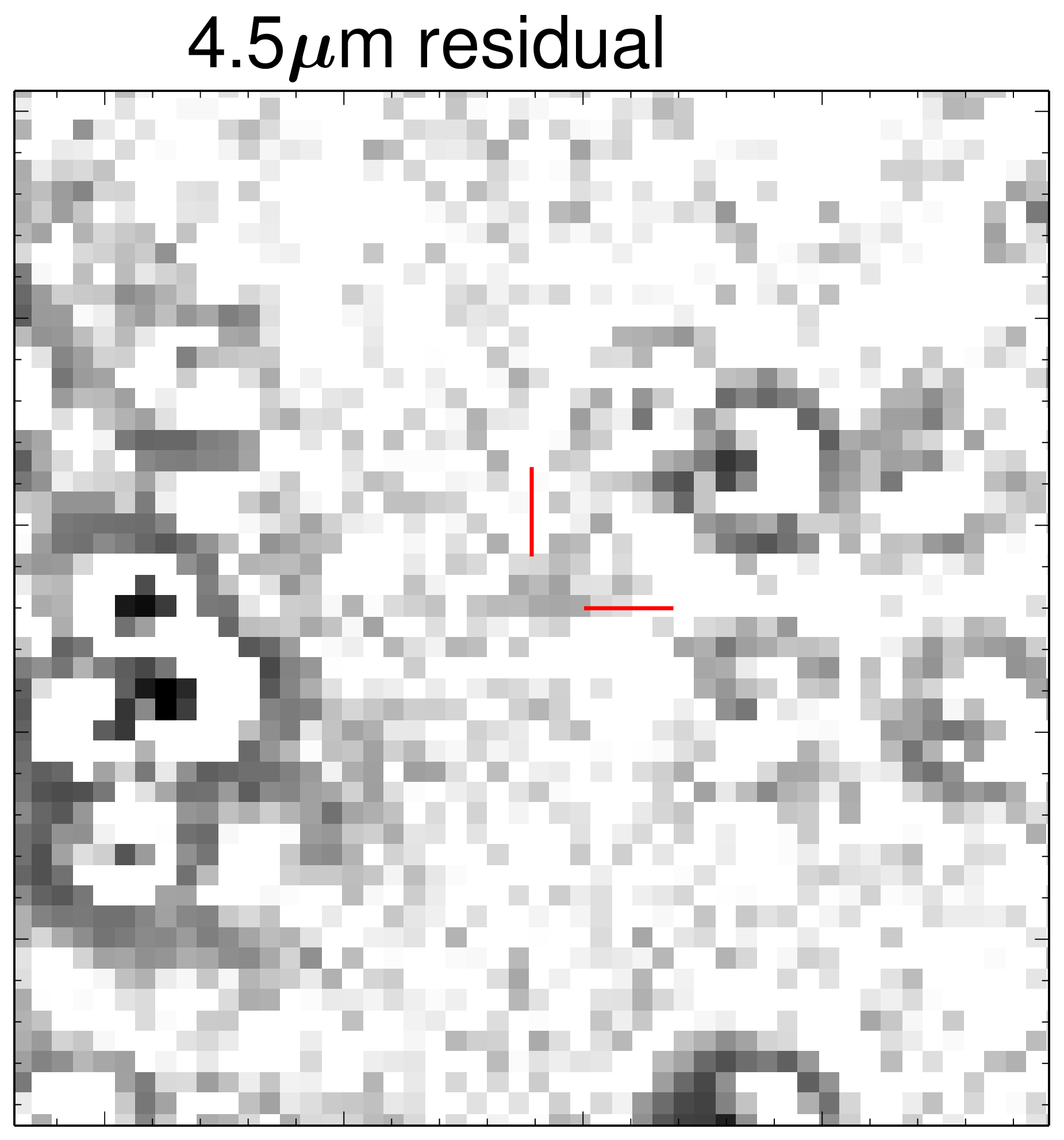}\\
\end{tabular}
\end{minipage}
\end{center}
\caption{Object MACS1149-JD from \citet{zheng12} shown in combined WFC3-IR colors ({\bf left}),  ${\chone}$  ({\bf middle}) and ${\chtwo}$ ({\bf right}) in $30\arcsec \times 30\arcsec$ boxes. Bottom row shows IRAC residuals  using {\tt TFIT} \citep{tfit} after subtracting all nearby objects detected in F160W band (excluding the main object). When performing photometry, all objects (including the main object) are fit simultaneously. North is up and East is left; $30\arcsec$
    corresponds to $\sim 200\mbox{kpc}$ at $z=9.5$ and magnification $\mu=14.5$.}
\label{fig:1149jd}
\end{figure}

In addition to the detection in 4.5 $\mu$m reported in
  \citet{zheng12}, we are also able to report a marginal detection of
  MACS1149-JD in 3.6 $\mu$m. We measure the IRAC fluxes using TFIT
  \citep{tfit}, which uses cutouts of each object in the
  high-resolution, e.g., F160W image, convolves them with PSF
  transformation kernels (from F160W to 3.6/4.5 $\mu$m) to prepare the
  low-resolution templates, and adjusts the normalization of each
  template to best match the surface brightness distribution of the
  IRAC images. Because of the large differences in angular resolution
  between HST and IRAC PSFs, we use the IRAC PSFs directly as the
  convolution kernels. To avoid the overcrowded region at cluster
  centers, we include only objects detected in F160W within a
  $20\arcsec \times 20\arcsec$ box centered at MACS1149-JD. To deal
  with local sky background, we measure the local sky level around
  MACS1149-JD within a $48\arcsec \times 48\arcsec$ box after masking
  out the detected sources. We subtract the median value of the sky
  pixels from the IRAC images and calculate the 1-$\sigma$ deviation
  as the sky level uncertainty. We then inflate the RMS image by the
  local sky uncertainty, and calculate the magnitude errors from the
  full covariance matrix of the templates included in the fit. The
  TFIT-measured fluxes represent the fluxes within the same
  isophotal aperture as in F160W (\texttt{MAG\_ISO} reported from
  SExtractor). Finally, we apply an aperture correction of $-0.4$ mag
  to match our \texttt{MAG\_ISO} in F160W to the reported total F160W magnitude
  from \citet{zheng12}. The IRAC magnitudes measured this way (also
  listed in Table~\ref{tab:1149}) are $[\chone] = 25.7 \pm 0.5$ mag and
    $[\chtwo] = 25.0 \pm 0.2$ mag, which are in agreement with \citet{zheng12}. We also list in Table~\ref{tab:1149} the magnitude errors if the RMS
  images were not inflated by sky level uncertainty, and clearly local
  sky uncertainty dominates the errors reported by the RMS image
  alone.  

\begin{figure}[h]
\begin{center}
\includegraphics[width=0.5\textwidth]{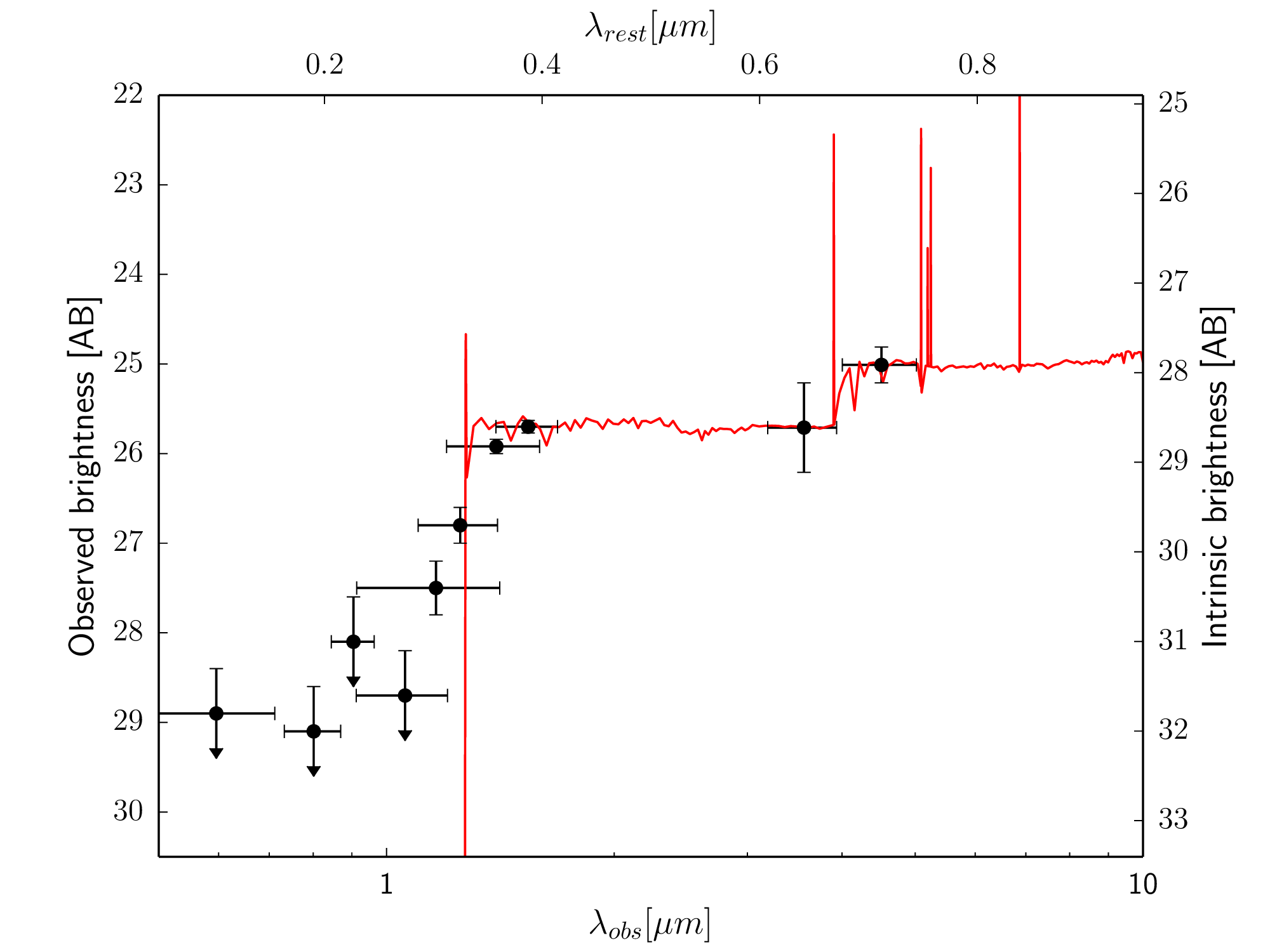}
\end{center}
\caption{SED fit for $z=9.5$ MACS1149-JD candidate behind  MACSJ1149.5$+$2223 from \protect{\citet{zheng12}}. Here the points show the observed photometry from
{\hst}/SST (the upper limits are 1-$\sigma$), and the line is the best-fit model from Le Phare
(including emission lines). On the right vertical axis, we show the intrinsic magnitudes corrected using magnification $\mu=14.5$.}
\label{fig:sed1149}
\end{figure}

\begin{figure}[hb]
\begin{center}
\includegraphics[width=0.5\textwidth]{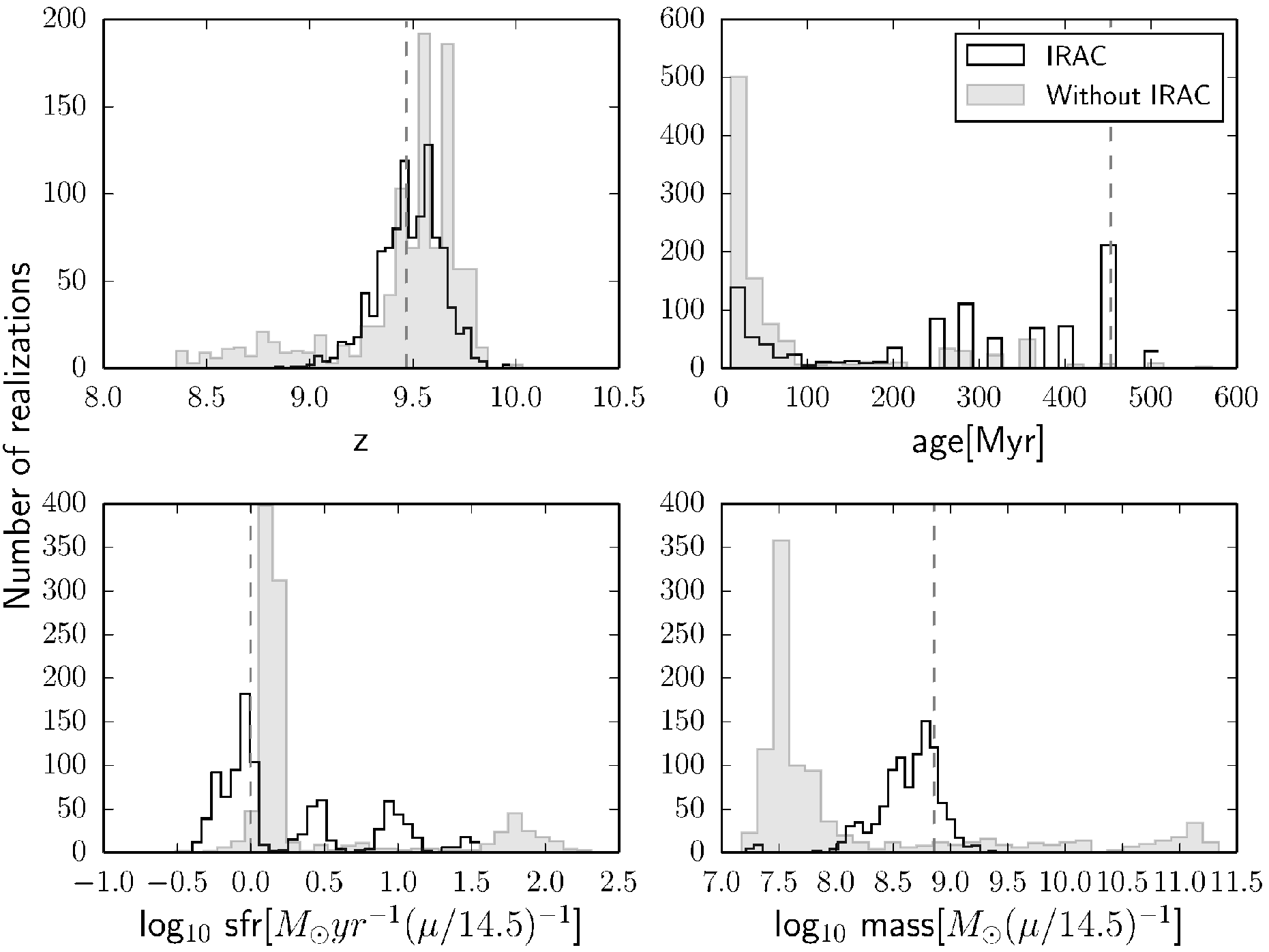}
\end{center}
\caption{Stellar population parameters for $z=9.5$ MACS1149-JD candidate behind  MACSJ1149.5$+$2223 from \protect{\citet{zheng12}}. To estimate the uncertainty on these
parameters, we use a simple Monte Carlo simulation as described in \protect{\citet{surfsup2}}. The open histogram shows results as derived upon the inclusion of {\surfsup} photometry (for the {\hst} we use data from \protect{\citealp{zheng12}}). The best fit values are given by the vertical dashed lines (see also Table~\ref{tab:1149}). To illustrate the importance of the IRAC data in modeling
these galaxies, the results without the IRAC data are indicated by shaded histograms,
slightly offset for clarity. While the photometric redshifts are robust to the exclusion
of the IRAC data, the SFRs, stellar masses and ages are not.}
\label{fig:stellar1149}
\end{figure}

  After performing IRAC photometry, we then perform SED fitting (see
  Fig.~\ref{fig:sed1149}) using {\tt LePhare} \citep{ilbert06,
    ilbert09, arnouts99} including fluxes from all available $HST$ and
  $Spitzer$ filters. The templates we use are from \citet{bc03}, but
  we also add a contribution from nebular emission lines to the
  templates (see \citealp{surfsup2} for details). This is especially
  important for an accurate measurement of the
  SFR and stellar masses using {\Spitzer} bands \citep{smit13}. We
  estimate that MACS1149-JD has a stellar mass of $M^* =
  7^{+1}_{-5}\times 10^{8} M_{\odot}$ (corrected for lensing using
  magnification $\mu = 14.5^{+4.2}_{-1.0}$ from \citealp{zheng12}) and
  an age of $\sim 450 \mbox{Myr}$. We report here the best fit
  parameters, and the uncertainties which are calculated from the Monte
  Carlo samples using the methodology described in detail
  \citet{surfsup2}. We estimate the errors by
  calculating the RMS of the samples. Full results are reported in
  Table~\ref{tab:1149}.  In Figure~\ref{fig:stellar1149} we show the
  marginalized probabilities for stellar population parameters. To
  illustrate the importance of the IRAC data in modeling these
  galaxies, we show the results without and with the IRAC data. While
  the photometric redshifts are robust to the exclusion of the IRAC
  data, the SFR and stellar mass and ages are not, clearly showing the
  importance to adding IRAC data. 

  The fitting results from the full photometry (see
  Table~\ref{tab:1149}) are broadly in agreement with those best-fit
  values derived by \citet{zheng12}. The one possible exception is the
  mean luminosity-weighted stellar age of MACS1149-JD, which is
  constrained in \citet{zheng12} to be younger than $275\mbox{Myr}$ at
  the 2-$\sigma$ level for the similar set of models that we employ
  here. In our fitting, both the best-fit model to the observed
  photometry and the models for more than half of our Monte-Carlo
  realizations have a mean luminosity-weighted stellar age in excess
  of the 2-$\sigma$ limit derived in \citet{zheng12}. The difference
  can perhaps be explained by the additional detection in the $\chone$
  band and the decreased uncertainty in the $\chtwo$ band detection,
  which allows for a more robust measure of the $4000\mbox{\AA}$
  break. As a result, there exists a hint from our data that
  MACS1149-JD contains an evolved stellar population\footnote{The age
    of the universe at $z\sim9.5$ is $\sim520\mbox{Myr}$.} for its
  redshift (i.e., $>$275 Myr), though we cannot definitively rule out
  younger ages.

\begin{deluxetable}{lc}
 \tablecolumns{2}
\tablewidth{0pc}
 \tablecaption{Properties of $z=9.5$ MACS1149-JD candidate behind  MACSJ1149.5$+$2223 from \protect{\citet{zheng12}}}
\startdata
\hline \hline
$[{\chone}]$ &  $25.7 \pm 0.5$ ($25.70 \pm  0.17  \pm             0.49$)\\
$[{\chtwo}]$ &  $25.0 \pm 0.2$ ($25.01 \pm 0.078 \pm 0.21$)  \tablenotemark{(a)}\\
\hline
F606W &    $<28.9$  \tablenotemark{(b)}  \\ 
F814W &    $<29.1$  \\
F850LP &    $<28.1$   \\   
F105W &    $<28.7$      \\ 
F110W &    $27.5 \pm 0.3$ \\ 
F125W &    $26.8 \pm 0.2$  \\
F140W &    $25.92 \pm 0.08$  \\
F160W & $ 25.70 \pm  0.07$\\
\hline
$z_{\rm phot}$ & $9.5 \pm 0.2$ \\
$SFR$ & $1.0^{+5.0}_{-0.4} (\mu/14.5)^{-1} M_{\odot} \mbox{yr}^{-1}$\\
$M^*$ & $7^{+1}_{-5}\times 10^{8} (\mu/14.5)^{-1} M_{\odot}$\\
Age & $450^{+30}_{-360} $\\
$\mu$ & $14.5^{+4.2}_{-1.0}$\tablenotemark{(c)}\\
$[{\chone}]_{\rm int}$ & $28.6^{+0.9}_{-0.8}$ \\
$[{\chtwo}]_{\rm int}$ & $27.9^{+0.6}_{-0.4}$\\
\enddata
\tablenotetext{(a)}{When estimating the magnitude errors we include both the contribution from the statistical error and systematic error due to the uncertainties in the local background error (see Sect.~\ref{sec:1149}). In parenthesis we list these two contributions separately.} 
\tablenotetext{(b)}{For F160W and all bluer bands we use {\hst} photometry from \citet{zheng12}. We matched in aperture our measured {\Spitzer} magnitudes, ensuring that colors are measured accurately. For non-detections 1-$\sigma$ detection limits are given.}
\tablenotetext{(c)}{\citet{zheng12}}
\label{tab:1149}
\end{deluxetable}

 \subsection{Prospects for Atacama Large Millimeter Array (ALMA) Followup}\label{sec:alma}
 While {\Spitzer} data increase our confidence in photometric redshift
 determination of $z\sim 7$ sources, the ultimate confirmation will
 come from spectroscopy. Spectroscopy is hard to do for typically
 faint high redshift sources, and it is thus an area where
 gravitational lensing magnification helps greatly (e.g.,
 \citealp{schenker12,bradac12}). However, despite the magnification,
 spectroscopic redshifts have been measured for only a handful of
 sources close to the reionization epoch. The non-detections are
 interpreted as evidence for the increase in opacity of the
 intergalactic medium above $z\sim 6$ (\citealp{fontana10, vanzella11,
   pentericci11, ono12, schenker12, treu12, treu13, finkelstein13}; if
 one assumes no evolution in escape fraction and clumping factors
 between $z\sim 6$ and 7).

 High ionization and atomic fine structure lines are an alternative
 way to observe these high-redshift galaxies.  Strong C III] emission
 (rest-frame $\lambda = 1909\mbox{\AA}$) is seen in every single
 lensed galaxy spectrum at $z\sim 2$ with stellar masses $\lesssim
 10^{9}M_{\sun}$ and low metallicities (Dan Stark, private
 communication, see also \citealp{erb06}). It is expected that these
 lines will be present at higher redshifts as well. Another
 possibility is the {\ct} line (rest-frame $158\mu\mbox{m}$). It is
 the strongest line in star forming galaxies at radio through FIR
 wavelengths and much stronger than the CO(1-0) line (see
 \citealp{carilli13} for a review).  By observing {\ct} emission in
 $z\sim 7$ galaxies we would not only measure their redshift, but also
 probe the photodissociation region surrounding star forming regions
 \citep{sargsyan12}.  As noted by \citet{carilli13}, the
 interpretation of {\ct} emission is not straightforward, because
 {\ct} traces both the neutral and the ionized medium and it appears
 to be suppressed in high density regions. Despite these difficulties,
 however, the {\ct} line is proving to be a unique tracer of galaxy
 dynamics in the early universe (see \citealp{carilli13} for an
 excellent compilation of results and references therein).

 Using the sample from the first {\surfsup} cluster, we now attempt to
 predict the rest-frame far infra-red (FIR) luminosity at $z\sim 7$
 and the expected {\ct} flux (for MACS1149-zD the line is
 unfortunately outside the current ALMA frequency range). We start by
 using the {\it lensed} (observed) infra-red luminosity predicted from
 SED fitting using {\tt LePhare} \citep{ilbert06, ilbert09, arnouts99}
 of the brightest z-band dropout from \citet{hall12}. The extrapolated
 IR luminosity for object \#3 is $L^{\mbox{lensed}}_{\rm FIR} = \mu
 L_{\rm FIR}= 1.2^{+2.4}_{-0.8}\times 10^{12}{L_{\sun}}$ (where $\mu$
 is the magnification; $\mu = 12 \pm 4$. Note that there are different
 definitions of FIR in the literature, for the purpose of this
 estimate $L^{\mbox{lensed}}_{\rm FIR}$ is defined as integrated
 luminosity from $8-1000\mu\mbox{m}$. One caveat is that we determine
 this luminosity by extrapolating the SED, hence the estimates are
 highly uncertain. To determine $L_{\mbox{\ct}}$ we use the
 $L_{\mbox{\ct}}/L_{\rm FIR}$ luminosity ratio from
 \citet{wagg12}. These authors find that the {\ct}/FIR luminosity
 ratio at high redshift is $8\times 10^{-4}$, which is lower than that
 of the Milky Way; $3\times 10^{-3}$ \citep{carilli13}. Hence we
 (conservatively) adopt the former. This suggests the {\ct} line
 luminosity of $L_{\mbox{\ct}} \simeq 10^9{L_{\sun}}$ and translates
 into a velocity integrated flux of $S_{\mbox{\ct}}\Delta v \lesssim 1
 \mbox{Jy km\,s}^{-1}$.  Such fluxes are easily reachable with
 ALMA. We caution, however, that this is a rough estimate, as $L^{\rm
   lensed}_{\rm FIR}$ and the $L_{\mbox{\ct}}/L_{\rm FIR}$ luminosity
 ratio are all very uncertain.  Note that the approach we use to
 estimate flux is different from that used in \citet{surfsup2},
 however both yield consistent results. ALMA observations will test
 these assumptions, and {\Spitzer} data will allow for an efficient selection of
 sources that will likely show [CII] emission due to a presence of
 evolved stellar population.

\section{Conclusions} \label{sec:conclusions} {\surfsup} will produce
a major advance in our understanding of the formation of the first
galaxies, in particular regarding their star formation history and stellar 
properties. This program will enable us to probe smaller stellar masses ($\sim 10^8 M_{\odot}$) and specific
star formation rates ($\sim 10^{-9} \mbox{ yr}^{-1}$) at the highest
redshifts $z \gtrsim 7$. If these high-redshift galaxies are responsible for
reionization, they need to produce a sufficient number of Lyman-continuum photons in a
sustained way. Once these galaxies are identified, the IGM-ionizing photon flux will be estimated from the
star formation rate density, which will include contributions from  instantaneous star formation rate dominated by younger stars and the integrated rate given by the older population \citep{robertson13}.

In this paper and in \citet{surfsup2} we have demonstrated the
importance of using IRAC data to estimate stellar masses, ages, and
SFRs for $z\gtrsim 7$ galaxies. In particular, we have shown that
without IRAC data the stellar properties are not robustly
determined. At $z \sim 7$,
the addition of IRAC photometry in SED fitting significantly reduces
the biases in the estimated galaxy properties compared to using HST
photometry alone  \citep{surfsup2}. At $z \sim 9$, the lack of IRAC
photometry in SED fitting can even lead to an order-of-magnitude bias
in stellar mass, SFR and age estimates. Hence,
{\surfsup} will contribute significantly to accurate measurements of
the stellar mass properties for these galaxies and thus, help
constrain the IGM-ionizing photon flux.

Not only do we have a limited knowledge of the earliest formation of
galaxies, but our picture of galaxy formation at later times is also
lacking many details. The magnifying power of galaxy clusters also allows us to
explore otherwise unreachable populations at ``intermediate''
redshifts ($1<z<7$). We will be able to probe the conditions in typical
low-mass, star-forming galaxies at an at an epoch when they
are otherwise inaccessible. The magnified galaxies provide excellent
targets for exploiting the unique capabilities of new facilities like
ALMA and JWST. By studying galaxy clusters, {\surfsup} will also enable measurements of the
stellar mass function of $z\sim 0.3-0.7$ galaxy cluster members. The
survey reaches depths of $<10^8$ M$_{\odot}$ (or $<0.005 L^*$) for an
elliptical galaxy at $z = 0.7$ (our highest redshift clusters). The
large FOV of {\Spitzer} will allow us to study cluster members out to
$R_{\rm vir}$.

Finally, {\surfsup} will be a resource for the broader community for the
study of distant, magnified sources and IR properties of lower
redshift galaxies. Data for 9 out of 10 clusters have been taken. We
have made available high-level science products (mosaics and empirical PSF
measurements) for the Bullet Cluster and we plan on releasing all the
data in the near future.

\acknowledgements We would like to sincerely thank the anonymous 
referee for his extremely fast, professional, and thorough work. We
would also like to thank Benjamin Cl\'{e}ment for making the VLT
HAWK-I $\mbox{K}_{\rm s}$ band data for the Bullet Cluster available
to us. We would also like to thank Gabriel Brammer for extensive help
with Eazy and Mariska Kriek in addition for making their codes ({\tt
  Eazy} and {\tt FAST}) and tools publicly available. We would also
like to thank Diego Garcia Appaddoo for numerous discussions about
ALMA capabilities.  Observations were carried out using {\Spitzer}
Space Telescope, which is operated by the Jet Propulsion Laboratory,
California Institute of Technology under a contract with NASA. Also
based on observations made with the NASA/ESA Hubble Space Telescope,
obtained at the Space Telescope Science Institute, which is operated
by the Association of Universities for Research in Astronomy, Inc.,
under NASA contract NAS 5-26555 and NNX08AD79G and ESO-VLT telescopes.
Support for this work was provided by NASA through an award issued by
JPL/Caltech. Support for this work was also provided by NASA through
{\hst}-GO-10200, {\hst}-GO-10863, and {\hst}-GO-11099 from STScI. TS
acknowledges support from the German Federal Ministry of Economics and
Technology (BMWi) provided through DLR under project 50 OR 1308. TT
acknowledges support by the Packard Fellowship. HH is supported by the
Marie Curie IOF 252760, by a CITA National Fellowship, and the DFG
grant Hi 1495/2-1. SA and AvdL acknowledge support by the
U.S. Department of Energy under contract number DE-AC02-76SF00515 and
by the Dark Cosmology Centre which is funded by the Danish National
Research Foundation.  Part of the work was carried out by MB while
visiting Joint ALMA Observatory (JAO); MB acknowledges support for the
visit by JAO through ALMA visitor programe. Part of the work was also
carried out by MB and TT while attending the program ``First Galaxies
and Faint Dwarfs'' at KITP which is supported in part by the NSF under
Grant No. NSF PHY11-25915.  

{\it Facilities:} \facility{Spitzer
  (IRAC)}, \facility{HST (ACS/WFC3)}, \facility{VLT:Yepun (HAWK-I)}

\bibliography{bibliogr_clusters,bibliogr_gglensing,bibliogr,bibliogr_cv,bibliogr_highz}
\bibliographystyle{apj}

\end{document}